\documentclass[11pt,letterpaper]{article}
\usepackage{graphicx} % Required for inserting images
\usepackage{booktabs}
\usepackage{tabularx}
\usepackage{amsmath,amssymb,amsthm}
\usepackage{amssymb}
\usepackage{stmaryrd}
\usepackage{braket}
\usepackage{quantikz}
\usepackage{adjustbox}
\usepackage[utf8]{inputenc}
\usepackage{newunicodechar}
\newunicodechar{ℕ}{\mathbb{N}}
\newunicodechar{·}{\cdot}
\usepackage[margin=1in]{geometry}
\usepackage{mathpazo}
\usepackage{dsfont}
\usepackage[hang,flushmargin]{footmisc}
\usepackage{hyperref}
\usepackage{algorithm}
\usepackage{algpseudocode}
\hypersetup{colorlinks=true, linkcolor=blue, urlcolor=blue, citecolor=Magenta}
\newcommand{\forshorai}{\url{https://forshor.ai}}
\newcommand{\verifiedqc}{\url{https://verifiedqc.github.io/ForShor/}}
\usepackage{caption}
\theoremstyle{plain}                     
\newtheorem{theorem}{Theorem}[section]
\newtheorem{lemma}[theorem]{Lemma}
\newtheorem{corollary}[theorem]{Corollary}
\newtheorem{definition}[theorem]{Definition}
\newtheorem{remark}[theorem]{Remark}

\usepackage{listings}
\usepackage[dvipsnames]{xcolor}
\lstdefinelanguage{Lean}{
  morekeywords={def,abbrev,structure,inductive,where,fun,match,with,if,then,else,
                some,none,by,let,theorem,lemma,example,namespace,end,open,deriving},
  sensitive=true,
  morecomment=[l]{-},
  morecomment=[s]{/-}{-/},
  morestring=[b]",
  literate=
    {ℕ}{{$\mathbb{N}$}}1
    {ℤ}{{$\mathbb{Z}$}}1
    {→}{{$\to$}}1
    {σ}{{$\sigma$}}1
    {≠}{{$\neq$}}1
    {×}{{$\times$}}1
    {·}{{$\cdot$}}1
}
\title{\bf Formally verified harness for Shor's order finding algorithm with $O(n^{2+\epsilon})$ gate cost}
\author{
  Anirudh Suresh\thanks{University of Maryland} \\
  \and
  Joshua M. Courtney\thanks{Zapata Quantum} \\
  \and
  Runzhou Tao\thanks{University of Maryland}\\
  \and
  Yudong Cao\thanks{Zapata Quantum. Corresponding author: \texttt{ycao@zapataquantum.com}.}\\
}
\date{\today}
\usepackage[style=numeric-comp,sorting=none,maxbibnames=99,backend=biber]{biblatex}
\begin{document}

\maketitle

\begin{abstract}
    It is commonly understood that factoring a general $n$-bit biprime via Shor's order finding algorithm requires $\widetilde{O}(n^3)$ gates by leveraging schoolbook integer arithmetics for addition and multiplication. Recent developments in quantum arithmetic circuits leveraging recursive Toom-Cook schemes have improved the cost scaling to $O(n^{2+\epsilon})$. Here $\epsilon>0$ depends on algorithm implementation details of the recursive multiplication steps, which allow for a vast number of design choices. The challenge in navigating such design space also presents an opportunity for demonstrating innovative practices in quantum software engineering in the era of open agentic science. We present a formally verified harness that ingests a design specification (namely a ``policy" that describes a recursive circuit design scheme) from an external source, which may be an AI agent, and produces a set of formalized proofs covering mathematical correctness of the algorithm implementation, asymptotic scaling of the end-to-end gate cost, as well as the correctness of translating the verified algorithm to a detailed circuit implementation with an intermediate representation. The harness is published as an open challenge where human researchers and AI agents can search for more efficient factoring circuits while retaining machine-checked guarantees of correctness and end-to-end asymptotic complexity.
\end{abstract}

\begin{figure}
    \centering
    \includegraphics[width=1\linewidth]{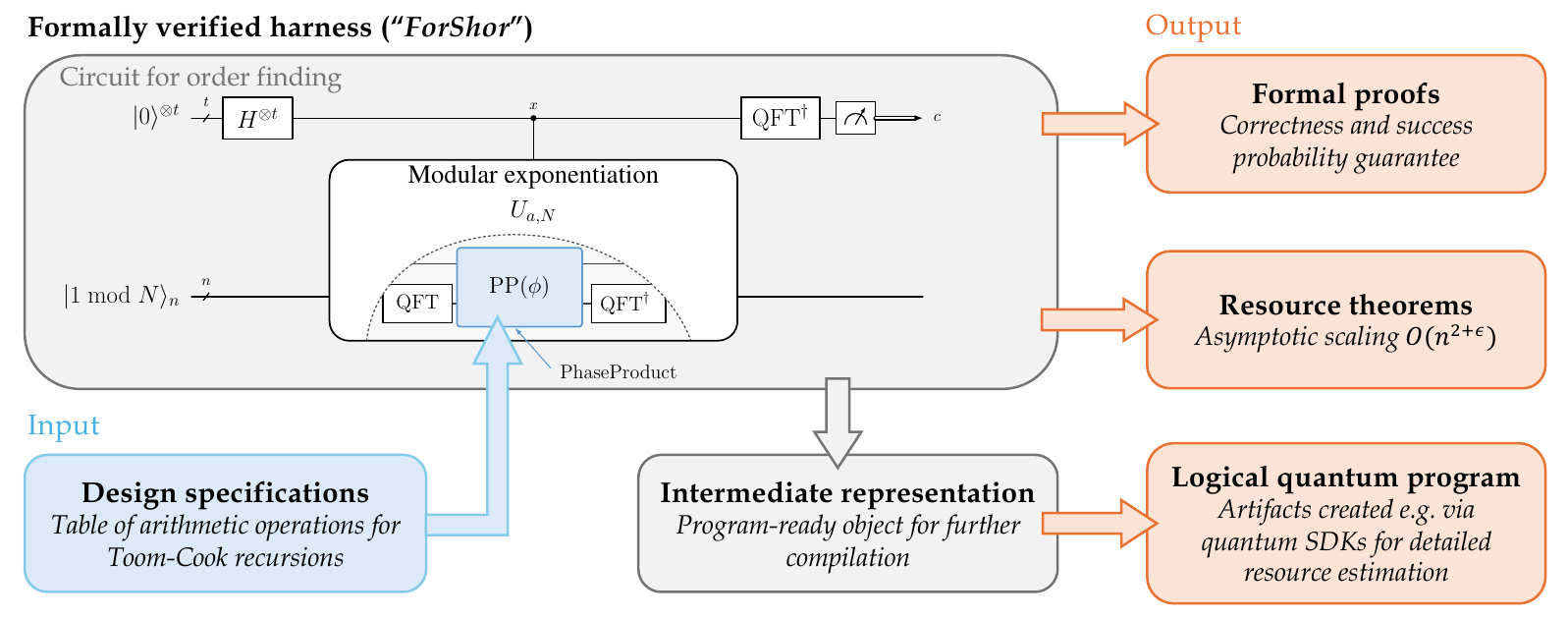}
    \caption{Schematic overview of the formally verified harness setup. A contributor supplies a design specification of the arithmetic circuit (namely a ``policy", see Section \ref{sec:PhasePolicyIsDifficult}), while the harness checks it in Lean, builds and lowers the order-finding circuit, and emits it to Qualtran.}
    \label{fig:forshor}
\end{figure}

\section{Introduction}

Integer factoring is a central use case for quantum computing, with consequences for both the study of quantum algorithms and the security of widely used public-key cryptography. Shor's algorithm gives a polynomial-time method~\cite{shor1997}, but the resources needed at practical input sizes depend strongly on how its arithmetic is implemented. Mapping this resource frontier helps assess the capabilities of future quantum computers and the risks to cryptographic systems. A practical reference point is the Gidney-Eker\aa{} construction~\cite{gidney2021factoring}, whose reported Toffoli count for an $n$-bit RSA integer scales as $\widetilde O(n^3)$ with schoolbook arithmetic, where $\widetilde O$ suppresses polylogarithmic factors. In 2024, Kahanamoku-Meyer and Yao (\cite{kahanamokuMeyer2024}, hereafter KMY) developed a phase-based multiplication scheme built on recursive
Toom-Cook multiplication~\cite[Sec.~1.3.3]{brentZimmermann2010} that yields an $O(n^{2+\epsilon})$ logical gate count for Shor's algorithm, for every fixed $\epsilon>0$ with suitable construction parameters. This gap in asymptotic scaling raises a practical question of whether the KMY construction offers a gate count advantage at relevant input sizes, and if so, where the crossover may be. Answering this requires explicit circuits and resource estimates under a common cost model~\cite{cao2025detailed}. We focus on gate count, since the construction studied here does not improve on the $n/2+o(n)$ logical qubit count already achieved for RSA factoring using short discrete log methods~\cite{ekera2017short,may2022compression,chevignard2024reducing,pappalardo2026shor}. Implementing and optimizing the recursive structure of the KMY construction raises two challenges:
\begin{enumerate}
    \item \textbf{Searching the circuit design space.} Breaking a multiplication into smaller ones and recombining their results can lower the total cost, but more pieces or costlier recombination can erase the gain. Choosing how to divide and recombine so that the circuit uses the fewest gates is the first challenge and an opportunity for AI-assisted research. Open projects~\cite{ecdsafailChallenge,beyondNLogN} show how public evaluation harnesses can autonomously process contributions from humans and agents. However, such open-ended agentic framework also raises a question about human understanding. A result may be technically correct yet hard for a human expert to parse, leading to the risk that autoresearch-generated and autoformalized results may accumulate faster than researchers can explain and absorb them~\cite[Secs.~6-7]{tao2026mathematics}. Inspecting and simplifying generated programs can reveal mathematical structure that guides further improvements~\cite{romeraparedes2024mathematical}. In quantum circuit search, structured programs can likewise help researchers identify reusable ideas, such as recovering branch decisions from the arithmetic state to reduce stored history~\cite{gu2026circuits}. These successes raise the question of how useful insights can be reliably drawn from many unfamiliar submissions, and how much researcher effort reusing them will take. The challenge is to give humans and agents enough freedom to discover useful designs while structuring submissions so that researchers can readily inspect, compare, explain, and build upon them.
    
    \item \textbf{Certifying the resulting circuits.} Each design needs correctness and gate-count guarantees that extend beyond tested instances. Tools such as Qualtran~\cite{harrigan2024expressing} and Qiskit~\cite{javadiAbhari2024qiskit} support circuit construction and analysis, but hardware limits and the exponential cost of general state-vector simulation restrict testing. Formal proofs must account for register sizes, ancilla uncomputation, relative phases, and approximation errors, with gate counts tied to the implemented operations under an explicit cost model. Existing methods support compositional reasoning about quantum programs~\cite{rand2018qwire,liu2019qhl,hietala2021programs,chareton2021qbricks} and semantics-preserving compilation and optimization~\cite{hietala2021voqc,li2022oracles}. Peng et al.~\cite{peng2023certified} have established correctness, success guarantees, certified gate counts, and executable circuit generation for Shor's algorithm. Recursive fast arithmetic adds a further requirement, namely that the surrounding proofs must remain reusable as arithmetic designs change, supporting a continuing search for improvements. Recent work likewise calls for machine-checked correctness and cost proofs for parameterized circuit families in AI-assisted quantum optimization~\cite{khattar2026verifiable, crichigno2026automating}.
\end{enumerate}

Together, these two challenges call for a close connection between the freedom to explore arithmetic designs and the ability to understand and certify them. Recent works in AI assisted mathematics~\cite{openai2026navierstokes,openai2026mathematics} and quantum circuit optimization~\cite{gu2026circuits, khattar2026verifiable, shirobokov2026discoveringformallyverifiedquantum,crichigno2026automatingquantumcircuitoptimization} gives this need added urgency. As these efforts produce more results and implementations, researchers need ways to inspect their structure, check their assumptions, and incorporate them into larger constructions. For fast quantum order finding, this means we need an interface between arithmetic design and verification that specifies which choices can vary and which constraints each proposed design must
satisfy. Formally proving that a new design satisfies those constraints should allow one to reuse the surrounding correctness proofs for the end-to-end implementation of the order finding circuit. The implementation can then be instantiated as concrete circuits for resource estimation.

We present \textbf{ForShor}, a formally verified harness for Shor's order finding algorithm for factoring using recursive quantum multiplication. The harness is published as an open challenge akin to the previously mentioned examples\footnote{The open challenge, submission interface and leaderboard are at \forshorai. Lean development and an interactive graph of declarations are at \verifiedqc.\label{fn:sites}}. However, the setting considered here is far more restrictive. Participants can only submit a \emph{policy} (Definition~\ref{def:policy}, also see Figure~\ref{fig:forshor}) that conforms to a set of requirements which guarantee that it legitimately specifies the design of KMY's recursive multiplication subroutine called \texttt{PhaseProduct}.
Any policy proved legitimate in Lean (Section~\ref{sec:PhasePolicyIsDifficult}) yields a correct order finding circuit, regardless of how the policy was produced.
Our proofs follow the hierarchy from factoring through order finding and modular exponentiation to arithmetic gates, establishing success guarantees, approximation bounds, circuit identities, and asymptotic gate counts.
They also certify the lowered circuit from which Qualtran bloqs are extracted for resource estimation.

As Section~\ref{sec:PhasePolicyIsDifficult} and Appendix~\ref{sec:InterpolationTableOpt} show, an exact algorithm exists for the optimum over independent recursion trees, a class that contains every policy, but its table-construction step is out of reach at cryptographic sizes, which opens the problem to heuristic search.
Optimizing \texttt{PhaseProduct} involves two linked choices at each recursion level: how many parts to split the register into (namely the \emph{arity}, denoted as $k$), and which operations to perform on those parts (the \emph{program}, denoted as $\Pi$). The exact algorithm constructs the cheapest table for every arity and width by exhaustive search and combines the results by a dynamic program over register width and leaf-rotation count.
Though it guarantees an optimal T-gate count, its table-construction step becomes intractable well before cryptographically relevant sizes: at arity 8 and $n=2048$ the third layer of its search graph alone holds $2.9\times10^{10}$ states ($7.5\times10^{3}$~GiB), and at the measured $540$~ns per candidate with $O(K^4)$ candidates per state, we estimate $3.2\times10^{51}$ core-hours to certify the best known tree (Appendix~\ref{sec:AppBProjection}).
With a large solution space, a clear objective and machine-checked acceptance criteria, the problem is a natural testbed for agentic optimization (Section~\ref{sec:PhasePolicyIsDifficult}).

In the remainder of this paper, Section~\ref{sec:PhasePolicyIsDifficult} states the policy search problem, which is the problem exposed to external participants. Section~\ref{sec:Harness} describes the implementation details of the harness, including its submission interface and its guarantees, and Section~\ref{sec:Qualtran} describes the emission to Qualtran for resource estimation.

% Appendix~\ref{sec:Background} reviews the algorithmic background and maps each component to the corresponding part of the formal development, and Appendix~\ref{sec:InterpolationTableOpt} gives the hardness and exact algorithm results for policy search.

\section{The \texorpdfstring{\texttt{PhaseProduct}}{PhaseProduct} Policy Search Problem}\label{sec:PhasePolicyIsDifficult}

Most of the computational cost in Shor's order-finding algorithm is in the modular exponentiation subroutine, being $O(n)$ controlled modular multiplications on an $n$-bit register (Appendix~\ref{sec:Background}).
Each multiplier and QFT is built from the \texttt{PhaseProduct} (\texttt{PP}) primitive~\cite{kahanamokuMeyer2024}, acting as $\ket{x}\ket{z}\mapsto e^{i\phi xz}\ket{x}\ket{z}$ on $w$-bit registers.
Splitting each register into $k\ge2$ chunks of width $h=\lfloor w/k\rfloor$ (the top chunk absorbs the remainder $r=w-kh$), $x=\sum_{j<k}2^{jh}x_j$ and $z=\sum_{j<k}2^{jh}z_j$, gives $x=X(2^h)$ and $z=Z(2^h)$ for the chunk polynomials $X(t)=\sum_j x_jt^j$ and $Z(t)=\sum_j z_jt^j$.
Their product $C(t)=X(t)Z(t)=\sum_{j=0}^{2k-2}c_jt^j$ has degree $2k-2$, so its $2k-1$ coefficients are fixed by its values at any $2k-1$ distinct points $p_0,\dots,p_{2k-2}$ (with $p=\infty$ standing for the leading coefficient $x_{k-1}z_{k-1}$). The coefficient vector $c=V^{-1}(C(p_\ell))_\ell =V^{-1}\bigl(X(p_\ell)Z(p_\ell)\bigr)_\ell$ for the Vandermonde interpolation matrix $V$ of the points with $V_{ij}=p_{i-1}^{j-1}$, $i,j\in\{1,2,...,2k-1\}$.
Substituting $t=2^h$,
\begin{equation}\label{eq:tc-identity}
\phi\,xz=\phi\,C(2^h)=\sum_{\ell=0}^{2k-2}\phi_\ell\,X(p_\ell)\,Z(p_\ell),
\qquad
\phi_\ell=\phi\sum_{j}2^{jh}\bigl(V^{-1}\bigr)_{j\ell}\quad, %\text{ (for $p_\ell=p$, $\phi_{(p)}:=\phi_\ell$)},
\end{equation}
so $e^{i\phi xz}$ factors into $2k-1$ \texttt{PhaseProduct}s with classically precomputed angles $\phi_\ell$. 
Each acts on evaluations only a few bits wider than $h$, formed in place from the chunks by reversible additions, negations and shifts and undone afterwards, at $O(w)$ cost, so the recursion $G(w)=(2k-1)\,G(w/k+O(1))+O(w)$ gives $O(w^{\log_k(2k-1)})$ (see Appendix~\ref{Background_FastPhaseProductConstruction}).

\begin{definition}[Interpolation table]\label{def:table}
An \emph{interpolation table} $T=(k,\mathbf{p},\Pi)$ consists of an arity
$k\ge 2$, an ordered list $\mathbf{p}=(p_0,\ldots,p_{2k-2})$ of
evaluation points in
$
\mathcal U
:= \mathbb Z
     \cup \{1/m : m\in\mathbb Z\setminus\{0\}\}
     \cup \{\infty\},
$
where evaluation at $\infty$ selects the leading coefficient, and a program $\Pi$ of arithmetic operations (including shifts, negations, and scaled
additions, as explained below), and $2k-1$ checkpoints on $k$ registers. The $\ell$-th
checkpoint applies the child \texttt{PhaseProduct} with angle $\phi_\ell$.
\end{definition}

\begin{definition}[Policy]\label{def:policy}
A \emph{policy} is an ordered list of pairs
$(\underline w_1,T^{(1)}),\ldots,(\underline w_m,T^{(m)})$, where each $T^{(j)}$ is an
interpolation table and the register-width thresholds satisfy
$\underline w_1>\cdots>\underline w_m$.
For a \texttt{PhaseProduct} on $w$-bit registers, the policy
selects the first table $T^{(j)}$ with $w\ge\underline w_j$.
If no such table exists, it selects the schoolbook construction.
\end{definition}

To define the basic arithmetic operations that constitute a program $\Pi$ in a table, write $r_0,\ldots,r_{k-1}$ for the current register values, each an
integer linear combination of the initial chunks $x_j$ or $z_j$. The arithmetic
instructions act as
\[
\begin{aligned}
\texttt{addScaled}(i,j,\eta,s)&:\quad r_i \leftarrow r_i+\eta\,2^s r_j,\\
\texttt{negate}(i)&:\quad r_i \leftarrow -r_i,\\
\texttt{shiftL}(i,s)&:\quad r_i \leftarrow 2^s r_i,\\
\texttt{shiftR}(i,s)&:\quad r_i \leftarrow r_i/2^s,
\end{aligned}
\]
where $i,j\in\{0,\ldots,k-1\}$, $i\ne j$ for scaled addition,
$\eta\in\{-1,+1\}$, and $s$ is a nonnegative integer.
Registers not targeted by an instruction remain unchanged.
A right shift is permitted only when every coefficient of the
symbolic expression for $r_i$ is divisible by $2^s$, ensuring exact
division for every input without discarding nonzero bits.
The program starts with $r_i=x_i$ and is applied identically to the
chunks of the other operand $z$. At each checkpoint, the child
\texttt{PhaseProduct} acts on the corresponding pair of registers,
changing only their phase and leaving their arithmetic values unchanged.
The register widths needed to implement these operations without
overflow are determined separately by the width scan.

\begin{figure}
    \centering
    \includegraphics[width=1\linewidth]{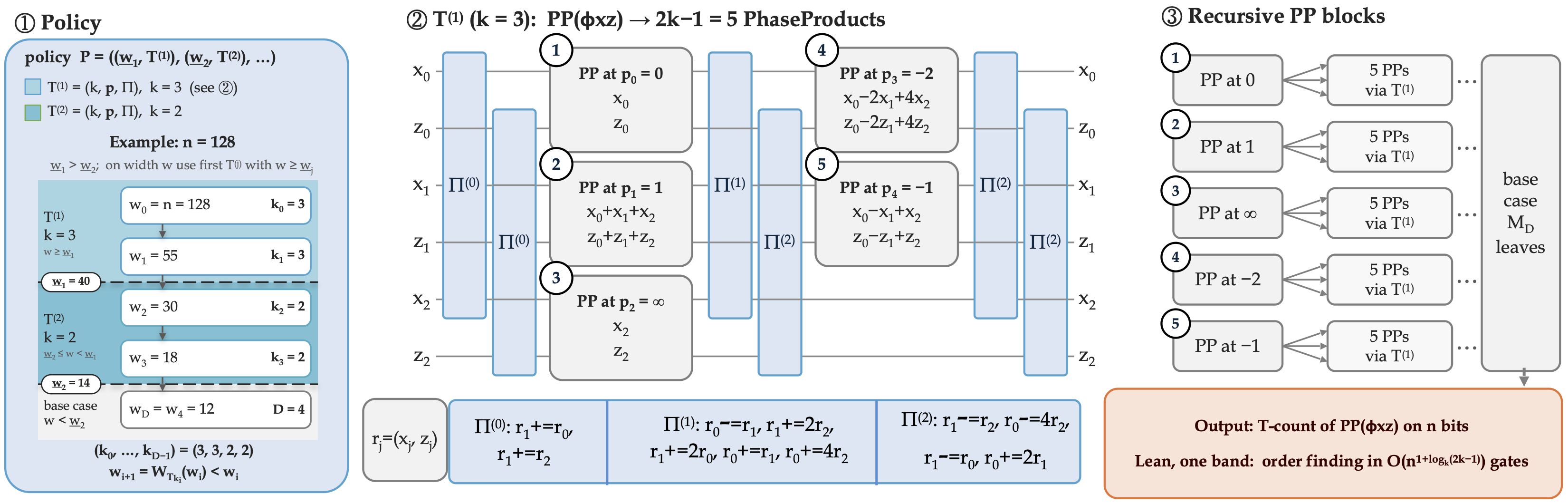}
    \caption{Schematic recursive \texttt{PhaseProduct} policy. 
    (1) An policy $P=((\underline w_1,T^{(1)}),(\underline w_2,T^{(2)}),\ldots)$ (Definition~\ref{def:policy}) pairs register width thresholds $\underline w_1>\underline w_2>\cdots$ with interpolation tables $T=(k,\mathbf p,\Pi)$ (Definition~\ref{def:table}), specifying one recursion level number of chunks (arity) $k$, ordered interpolation points $\mathbf p$ and a program $\Pi$ of scaled additions, negations, and shifts.
The first table decomposes a \texttt{PhaseProduct} on $w$-bit registers with $w\ge\underline w_j$, so widths $w_{i+1}=W_{T_{k_i}}(w_i)<w_i$ from $w_0=n$ induce an arity sequence $k_0,\ldots,k_{D-1}$ down to a base level $D$.
For $n=128$ as shown here, thresholds $(40,14)$ with tables of arity $3$ and $2$ (the latter from Appendix~\ref{Background_FastPhaseProductConstruction}) give an arity sequence $(3,3,2,2)$ and $D=4$.
(2) A Lean-verified $k=3$ table $T^{(1)}$ is decomposed with interpolation points $\mathbf p=(0,1,\infty,-2,-1)$. The table uses 11 scaled additions within program blocks $\Pi^{(0)},\Pi^{(1)},\Pi^{(2)}$.
(3) At each (numbered) checkpoint, a child \texttt{PhaseProduct} of width $w_{i+1}$ is instantiated and decomposed by the width-selected table, recursing until the $M_D$ base case leaves.
Total T-count sums arithmetic and leaf operations (Eq.~\eqref{eq:policyT}). 
The Lean gate-count bound (Theorem~\ref{thm:gatecount}) holds for every valid policy.}
    \label{fig:PhaseProduct_TableOpt}
\end{figure}

For example, consider the policy $((40,T^{(1)}),(14,T^{(2)}))$,
where $T^{(1)}$ and $T^{(2)}$ are interpolation tables of arity
$3$ and $2$, respectively. It selects $T^{(1)}$ for $w\ge40$,
$T^{(2)}$ for $14\le w<40$, and the schoolbook construction
for $w<14$. After selecting a table, the
implementation checks whether its decomposition reduces
the register width and falls back to schoolbook if it
does not.
For a given input width $n$, all \texttt{PhaseProduct} operations at
recursion depth $i$ share width $w_i$. A legitimate policy must ensure that each table satisfies
\begin{enumerate}
  \item[\textbf{C1}] there are exactly $2k-1$ points, one per coefficient of $C(t)=X(t)Z(t)$
  \item[\textbf{C2}] the points are distinct, so the interpolation matrix $V$ is invertible and~\eqref{eq:tc-identity} holds
  \item[\textbf{C3}] running the program from the initial state, each instruction ensures no right shift discards a nonzero bit, the $\ell$-th checkpoint finds its register holding exactly the evaluation row of $p_\ell$, and every point is consumed, in order
    \item[\textbf{C4}] the program returns every register to its initial value
\end{enumerate}
and that its thresholds satisfy $\underline w_1>\cdots>\underline w_m$. Thus, the recursive steps induce
an arity sequence $k_0,\ldots,k_{D-1}$, with $w_0=n$ and
$w_{i+1}<w_i$, terminating in the schoolbook construction
at recursion depth $D$. Figure~\ref{fig:PhaseProduct_TableOpt} shows a $k=3$ table and the recursion
it generates. In the harness implementation we require that these basic properties that make a policy legitimate be formally proven in Lean (Section~\ref{subsec:Submissions}).

% Formalization details are given on the \textbf{ForShor} website\footref{fn:sites}, and Appendix~\ref{Background_FastPhaseProductConstruction} explains why C1 and C2 are necessary. In short, C1-C2 make~\eqref{eq:tc-identity} valid, C3 makes the $\ell$-th checkpoint see $X(p_\ell)$ and $Z(p_\ell)$, so that the angle $\phi_\ell$ is correct, and C4 makes the circuit act only through a phase, namely diagonal in the computational basis $|x\rangle|z\rangle$. These are decidable propositions, established by a decision procedure where the kernel checks the submission's proofs of C1-C4 relative to standard logical axioms. In the case of the harness, a proof that relies on \texttt{sorry} (the placeholder in Lean for a missing proof) or that relies on an axiom introduced by the contributor is rejected.

We restrict attention to
policies that use at most one table $T_k$ for each arity
$2\le K$ for some maximum arity $K$ given a priori. For a table within a policy, for a given $k$ we fix the set $\mathcal{E}_k$ of $2k-1$ interpolation points to that of
KMY~\cite{kahanamokuMeyer2024}, so a table chooses only their traversal
order $\mathbf{p}$ within $\mathcal{E}_k$ and its program $\Pi$. 
We further take each child width $W_p$ to be the same given a parent width $w$.
We recognize this as a restricted family of policies, as opposed to the set of all valid policies. 
Even optimizing within this family, by numerical experiments we discover practical intractability for exact methods.
As such, the full policy optimization problem is at least as hard as the restricted policy search problem considered in this paper.
Each table $T$ associated with arity $k$ determines a \emph{child width function} denoted as $W_{T}(w)$, whose value is approximately $w/k$ plus ancillary bits accounting for register growth
during its program. Starting from $w_0=n$, a policy selects arity
$k_i$ at level $i$ and sets $w_{i+1}=W_{T_{k_i}}(w_i)$.
We require $w_{i+1}<w_i$ at every recursive level. At depth $D$,
the remaining products are computed by the schoolbook method,
with $O(w_D^2)$ operations per product.
As we detail below, we consider a restricted family of \texttt{PhaseProduct} schedules, retaining a constant $k$ at a given recursion level, with interpolation points defined from KMY~\cite[Appendix F.\ 2]{kahanamokuMeyer2024}. 

For each selected table $T_k$, define its \emph{arithmetic charge}
as $\chi_k:=A_+(T_k)+3A_-(T_k)+\nu(T_k)$, where $A_+$, $A_-$,
and $\nu$ count scaled additions, scaled subtractions, and explicit
negations, respectively, in the full program, including restoration.
The weights reflect the benchmark's negate-add-negate implementation
of subtraction. Shifts incur no direct arithmetic charge, and recursive
calls at interpolation points are costed separately. The charge weights and
width scan are detailed in Appendix~\ref{app:charge-width}.
Under the T-count model of Section~\ref{sec:Qualtran},
the policy has cost
\begin{equation}\label{eq:policyT}
\mathsf T_{\mathrm{PP}}=\sum_{i=0}^{D-1}
8\,M_i\,(w_{i+1}-1)\,\chi_{k_i}
+\mathrm{leaf}(M_D,w_D),
\qquad M_i=\prod_{j<i}(2k_j-1),
\end{equation}
where $M_0=1$ and $M_i$ is the number of \texttt{PhaseProduct}
instances at level $i$. Here, $\mathrm{leaf}(N,w)$ is the total
base-case T-count for $M$ schoolbook products of width $w$,
including the synthesis cost of $Mw^2$ rotations and their
associated logical ANDs (Eq.~\eqref{eq:TPP}).

The \emph{policy search problem} is the following constrained
optimization problem. Given an input register width $n$, a maximum
arity $K$, and the fixed interpolation point set for each
arity, find an admissible policy that minimizes the total T-gate
count in Eq.~\eqref{eq:policyT} for a \texttt{PhaseProduct} on
$n$-bit registers. The optimization jointly chooses
(i) which arities $k\in\{2,\ldots,K\}$ to use;
(ii) for each chosen arity, an ordering $\mathbf{p}_k$ of its fixed
point set and an admissible program $\Pi_k$, defining a table
$T_k=(k,\mathbf{p}_k,\Pi_k)$; and
(iii) the width thresholds and table assignments defining the
policy's bands, with schoolbook multiplication used when no band
applies. At most one table per arity may be selected here. 
Meanwhile, the exact algorithm of Appendix~\ref{sec:InterpolationTableOpt} relaxes this and lets every recursive call choose its own table, letting its optimum lower-bound the optimum of the problem stated in this section. 
The challenge as constructed does not require one table per arity.
The objective includes both the arithmetic charges at all recursive
levels and the total schoolbook leaf cost, rather than merely the
cost of an individual table. Thus, the policy search problem optimizes over a restricted class of the policies in Definition~\ref{def:policy}. We optimize over the larger class of independent recursion trees in Appendix~\ref{sec:InterpolationTableOpt}.

Why is this problem difficult? 
First, the space of programs is unbounded within a table, so a program can visit the points in any order, write to any of the $k$ registers, keep partial sums between points, and scale by arbitrary shifts. The cost of computing the evaluation at the next interpolation point depends on the values currently stored in the registers, including intermediate results retained from earlier computations.
Second, a table is evaluated by the total cost of the recursive circuit it induces, not merely by the length of its program. A program with fewer additions may produce wider inputs at its checkpoints, increasing the cost of the recursive child calls and their descendants. Thus, the cheapest policy need not use the shortest instruction lists.
Finally, changing a table affects the cost at every recursion level that uses it. Also, adding an input bit may increase the largest block width for one arity but leave it unchanged for another. These effects can make the cheapest arity switch back and forth as $n$ increases. For example, with $K=3$ the exact optimum over independent trees uses arity $2$ at $n=60$, arity $3$ at $n=61$, $62$ and $63$, arity $2$ at $n=64$, arity $3$ at $n=65$ and arity $2$ again at $n=66$.

We summarize each table by three integers: one measures the cost of its additions and negations, and the other two specify how many extra bits are needed to store intermediate results (Theorem~\ref{thm:profile}). We optimize in two stages, first finding the cheapest table for every arity and profile, followed by a dynamic program which chooses whether to stop the recursio, or to use a particular table at every call  (Appendix~\ref{sec:AppBEndtoEnd}).
The \emph{inner} layer finds the cheapest table at each arity and width profile. 
This is performed by exhaustively enumerating skeletons of charged instructions and resolving the free shift amounts exactly by integer feasibility.
The \emph{outer} layer chooses whether to stop or which table to use for each call. Since the cost of leaf rotation synthesis is nonlinear in their total number, it carries the total number of rotations as a state. 
A dynamic program will combine $2k-1$ children for each call and apply the price over register widths and number of rotations. 
We find this program to scale as $O(\kappa^{2}n^{6})$ arithmetic operations, polynomial in $n$, where $\kappa = \min (K,n)$.
Integer feasibility inside the inner layer is solved by the Lenstra--Kannan algorithm~\cite{kannan1987minkowski}.
We recover an exact, deterministic algorithm running in time $2^{O(\kappa^{4})}\operatorname{poly}(n)$ (Theorem~\ref{cor:recExact}), fixed-parameter in the arity bound and not practical for RSA-2048.
The exponential cost is in the inner layer, measured in Appendix~\ref{sec:AppBInnerOuter}: the search graph for one table is memory-bound, and the time per candidate, $540$~ns at arity $8$, converts its size into core-hours. We leave further optimization to future work.

% We find that a related problem is hard, but it is distinct from the T-count optimization problem. 
% If the point set is part of the input and tables are ranked by a structural score (minimal number of arithmetic instructions plus a fixed weight per layer of phase products), the decision problem is NP-hard already for $k=2$, since choosing the order of the points encodes a Hamiltonian cycle (Appendix~\ref{sec:AppBSuppliedPoints}). 
% The hard instances use point sets the harness does not currently admit, and ignore width entirely. 
% This highlights point ordering as carrying combinatorial hardness, though we keep structural optimization secondary to a T-optimal policy search. 

We implemented the planner of Appendix~\ref{sec:InterpolationTableOpt} in Fortran~95 (Appendix~\ref{sec:AppBEndtoEnd}), with the exact $K=2$ closed form and, for $K\ge3$, tables built by exhaustive search over the KMY points. Each returned tree and table is re-verified by an independent replay. At $n=2048$ it returns the exact $K=2$ optimum in $5.6$~s and the $K=3$ optimum over its tables in about four minutes.
The cost lies in table construction, where for one arity-8 table at $n=2048$ the third layer of the search graph holds $2.9\times10^{10}$ states ($7.5\times10^{3}$~GiB, about $470\times$ a 16~GiB machine), and certifying the best known $n=2048$ policy is estimated at $3.2\times10^{51}$ core-hours (sensitivity span $3.2\times10^{51}$ to $4.9\times10^{80}$) at a measured $540$~ns per candidate, against $3.2\times10^{12}$ core-hours for the arity-4 decision.
At $K=3$ the exact optimal trees reach depth $4$ at $n=1024$ and $n=2048$, and at $K=2$ depth $6$ at $n=2048$.
Checking candidate policies in Lean is cheap by contrast, and their T-count follows in closed form from~\eqref{eq:policyT}.
Since the exact method is out of reach at $n=2048$, but we have an explicit objective function and easy machine-checked acceptance criteria, \texttt{PhaseProduct} policy optimization makes a natural target for heuristic search, with \textbf{ForShor} serving as the referee in full algorithmic integration.

\section{The \texorpdfstring{\textbf{ForShor}}{ForShor} Harness}
\label{sec:Harness}

\begin{figure}
    \centering
    \includegraphics[width=\linewidth]{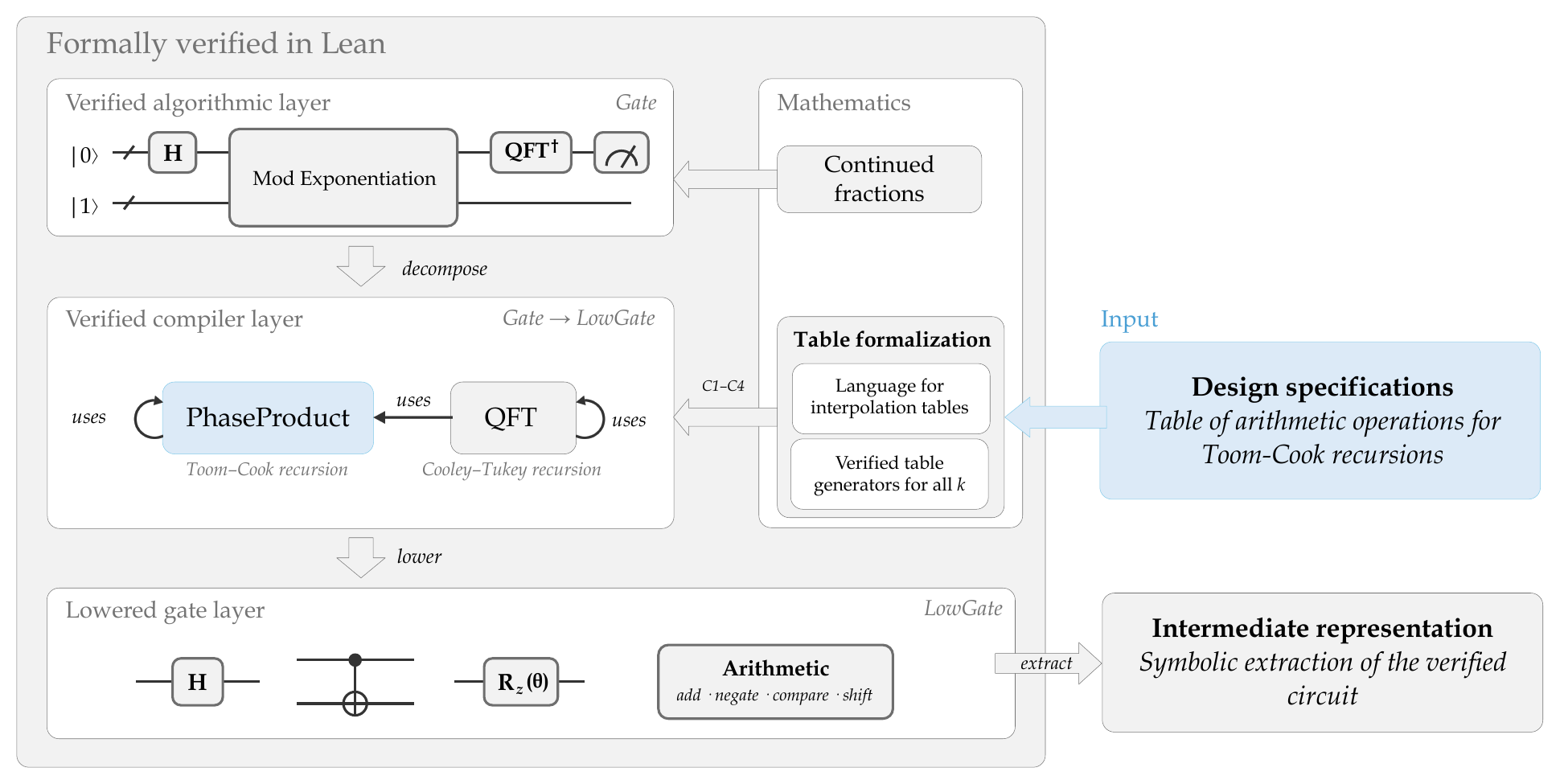}
    \caption{The verified part of Figure~\ref{fig:forshor} in detail. The order-finding circuit is written in \texttt{Gate}. Its \texttt{PhaseProduct} and QFT subroutines are constructed recursively once every table in the submitted policy passes C1--C4. A verified compiler produces the \texttt{LowGate} circuit from which the intermediate representation is extracted. Figure~\ref{fig:PhaseProduct_TableOpt} is an expanded view of the \texttt{PhaseProduct} part of this schematic.}
    \label{fig:harness}
\end{figure}

\subsection{Submissions}
\label{subsec:Submissions}

\textbf{ForShor} separates the development into a \emph{fixed harness} containing the circuit languages, the correctness criterion, the construction of the order finding circuit and all related theorems, and the emission pipeline. The {policy} (Definition~\ref{def:policy}) for recursion is the only part a participant provides (Figure~\ref{fig:harness}).\footnote{Submissions are made through \forshorai. A template \texttt{ShorPolicySubmission} file and the full Lean statements of C1--C4 are linked from \verifiedqc.} The harness decides whether submitted policies are valid, and if so the correctness and cost theorems, proved once for all valid policies, apply to the resulting circuit with no further proof.

% The design parameters form a \emph{policy} (Definition~\ref{def:policy}), where each band's threshold is the smallest register width at which its table is used. At or above its threshold a table decomposes every \texttt{PhaseProduct} the construction meets, so a wide table serves the top of the recursion and narrower ones the levels below it; below the last threshold the harness uses the naive implementation (Appendix~\ref{Background_PhaseProduct}), as it does whenever a table would not shrink the width. A single table used at every width is the \emph{one-band} policy, so a single-table submission is the special case at threshold $0$. A table's program acts not on quantum registers but on a \emph{symbolic model} in which each register holds a linear combination of the chunks $x_0,\ldots,x_{k-1}$ as its \emph{coefficient vector}, and a checkpoint marks where a smaller \texttt{PhaseProduct} is inserted. This makes a table independent of register width, so it is checked once and used at every width and level at which its band is selected.

Each policy consists of a list of tables and their thresholds. In Lean, a valid table is formalized as a record \texttt{ShorLoweringSetup} holding $k$, the points, the program, and a proof of each of C1--C4. A submission is a list of such records with their thresholds, together with a proof that the thresholds strictly decrease:
\begin{lstlisting}[language=Lean]
structure ShorPolicySubmission where
  bands  : List (ℕ × ShorLoweringSetup)
  sorted : (bands.map Prod.fst).IsChain (· > ·)
\end{lstlisting}
A policy is \emph{valid} when it can be packaged appropriately, where every band's table satisfies C1--C4 and the thresholds strictly decrease.

\subsection{Guarantees for Every Submission}
\label{subsec:Guarantees}

From a valid policy $P$ and an order-finding instance $(N,a)$, the harness builds an order-finding circuit $\mathcal{C}_P(N,a)$. The following theorems are proved once for every valid policy.

\begin{theorem}[Correctness, generic in the policy]
\label{thm:correctness}
Let $P$ be any valid policy and $(N,a)$ any order-finding instance ($0<a<N$, $\gcd(a,N)=1$). Let $\mathcal{C}_P(N,a)$ be run at any precision $\eta \le \eta_N$, where $\eta_N$ is the threshold fixed by the harness, which costs $O(\log n)$ extra work qubits.
Then
\[
\Pr\bigl[\text{successful order recovery from } \mathcal{C}_P(N,a)\bigr]
\;\geq\;
\frac{\kappa}{2\ln\bigl(\lfloor \log_2 N\rfloor+1\bigr)},
\qquad
\kappa=\frac{1}{\pi^{2}}.
\]
\end{theorem}

The bound is half that of the ideal circuit with exact modular exponentiation. The fast QFT is exact and the lowering preserves the state, so the whole loss comes from the approximate multiplications (Appendix~\ref{Background_ModMul}). 
The prescribed precision is chosen so that their accumulated error costs at most half the ideal bound, requiring $O(\log n)$ extra work qubits(leaving the total $O(n)$ work qubits count unchanged). 

The second theorem bounds the size of the same circuit. The lowered gate count of $C_P(N,a)$ is the cost of the \texttt{LowGate} circuit under the harness's cost model, which charges each elementary gate as one and each arithmetic gate linearly in its register width.

\begin{theorem}[Gate count, generic in the policy]\label{thm:gatecount}
Let $P = ((w_1, T^{(1)}), \dots, (w_m, T^{(m)}))$ be any valid policy with $m \ge 1$, and let $k_1$ be the number of chunks of $T^{(1)}$. At every precision at which the work register has $O(n)$ qubits, the lowered gate count of $C_P(N,a)$ is $O\bigl(n^{1+\log_{k_1}(2k_1-1)}\bigr)$, where $n$ is the bit length of $N$. In particular this holds at the precision of Theorem~\ref{thm:correctness}.
\end{theorem}

\begin{corollary}
\label{cor:scaling}
For every $\epsilon > 0$ there is a $k$ such that every valid policy whose first table has $k$ chunks yields an order-finding circuit with $O(n^{2+\epsilon})$ gates.
\end{corollary}

The count is of the lowered circuit, so register extension, workspace management and the base case are all charged, and the prescribed precision adds only $O(\log n)$ qubits, leaving the exponent unaffected. The lower bands of a policy act only on instances of bounded width, so they change the constant and not the exponent. The bound is on the same lowered circuit as Theorem~\ref{thm:correctness}, under the same hypotheses. Section~\ref{sec:Qualtran} evaluates more concrete resource estimations of the emitted circuit in Qualtran for one-band and varying-$k$ policies.

\subsection{Proof Architecture}
\label{subsec:ProofArchitecture}

We now describe how the harness establishes these guarantees, and why the proofs apply to every valid submission. The development has three layers, as shown in Figure~\ref{fig:harness}: i) The \emph{algorithmic layer} is written in \texttt{Gate}, a high-level language in which the QFT and \texttt{PhaseProduct} are single primitives, the latter with the semantics $\ket{x}\ket{z}\mapsto e^{i\phi xz}\ket{x}\ket{z}$. Two circuits live here. The ideal order-finding circuit uses an exact modular-exponentiation oracle, and the success argument of Shor's algorithm, including the continued-fraction postprocessing, is proved for it once. The implemented circuit replaces the oracle by the fast modular multiplier of Appendix~\ref{Background_ModMul}, built from these primitives, and is proved to approximate the ideal one. Neither argument depends on how \texttt{PhaseProduct} is implemented.
ii) The \emph{compiler layer} lowers \texttt{Gate} to \texttt{LowGate}. This is where the recursive constructions live and where the submitted tables enter, and the compiler is itself proved correct, so every guarantee established at the \texttt{Gate} level transfers to the lowered circuit. iii) The \emph{lowered gate layer} is the \texttt{LowGate} circuit of elementary gates and width-indexed arithmetic. The gate-count theorem is proved against its cost model, and the intermediate representation of Section~\ref{sec:Qualtran} is extracted from it.

Recursive constructions live in the compiler. It replaces each \texttt{Gate}-level QFT by its Cooley-Tukey decomposition into two half-size QFTs joined by a \texttt{PhaseProduct} (Appendix~\ref{QFTBackground}), and each \texttt{PhaseProduct} by its Toom-Cook decomposition into $2k-1$ smaller instances surrounded by interpolation arithmetic, recursing until the direct implementation is used. Its correctness therefore rests on the correctness of these two recursive decompositions, each proved to have the semantics of the primitive it replaces.
Previous work by Nanniyur and Jayanti~\cite{nanniyur2026machine} worked to formalize classical Toom-Cook multiplication. 
We focus on formalizing quantum Toom-Cook from first-principles in Lean, as well as its integration in \texttt{PhaseProduct} and in Shor's order-finding.

% \begin{lemma}[Table compilation]\label{lem:compile}
% Let $T$ satisfy C1--C4, and let $x,z$ be registers holding signed integers $\bar x,\bar z$ with enough clean guard bits for the widths derived from $T$. Then the circuit $\mathrm{compile}_T(\phi,x,z)$ satisfies $\mathrm{compile}_T(\phi,x,z)\ket{b}=e^{i\phi\bar x\bar z}\ket{b}$ on every such basis state, and hence on their span.
% \end{lemma}
% The proof is an induction along the program. C3 says the symbolic run reaches the $\ell$-th checkpoint holding the evaluation row of $p_\ell$, so the child applies $e^{i\phi_\ell X(p_\ell)Z(p_\ell)}$. By C1--C2 these phases multiply to $e^{i\phi xz}$ via~\eqref{eq:tc-identity}. By C4 the registers end where they started. The argument uses nothing but C1--C4, so it covers tables the harness has never seen, and applying it at each level reduces the recursion to the verified schoolbook leaf.

\begin{theorem}[Compiler correctness]
\label{thm:lowering}
Let $P$ be any valid policy and $G$ a \texttt{Gate} circuit of the order-finding construction. For every state $\ket\psi$ whose ancilla qubits are clean,
\[
\operatorname{evalL}\bigl(\operatorname{lower}_P(G),\ket\psi\bigr)=\operatorname{eval}(G,\ket\psi).
\]
\end{theorem}
Here, $\operatorname{eval}$ and $\operatorname{evalL}$ are the evaluators for the \texttt{Gate} language and the \texttt{LowGate} language respectively. When a submitted policy, which enters the compiler(Figure~\ref{fig:harness}), reaches a \texttt{PhaseProduct}, it selects the band for the current register width and decomposes by that band's table, or uses the direct implementation if no band applies or the table does not reduce the width. Everything else is identical across submissions, so Theorem~\ref{thm:lowering}, and with it the theorems of Section~\ref{subsec:Guarantees}, holds for an arbitrary valid policy.

\section{Translation to Qualtran}
\label{sec:Qualtran}

The verified circuit lives in Lean, while resource estimation is performed in Qualtran~\cite{harrigan2024expressing}. 
Qualtran has previously been used for resource estimation, allowing for scalability through symbolic gate calculation, including for the Toom-Cook multiplication we consider here~\cite{cao2025detailed}.
We bridge the gap between Lean and Qualtran by exporting a lowered order-finding circuit from Lean as an intermediate representation. 
Since Lean is designed for constructing and checking proofs, not for large-scale computation, evaluating a circuit for order-finding of a 2048-bit number inside Lean would be extremely slow and memory-intensive. 
By exporting a compact symbolic description of the circuit, with recursive components such as the \texttt{PhaseProduct} and the QFT described by their recursive rule and auxiliary values like subproblem widths and phase coefficients exported as lookup tables, Qualtran can evaluate the description and perform logical resource estimation. 
The IR is extracted from the same Lean definitions as the correctness theorems, translated into a small language of primitive operations, register expressions, and recursive calls. 
Because the IR is derived from the verified definitions rather than from a separate implementation of the algorithm, the circuit that Qualtran prices is the circuit that was proved correct. 

The formal verification ends where Qualtran begins.
Once emitted, the IR is read by dictionaries that map tags, vocabulary, and operator names to a hard-coded analog in Qualtran.
Lean-centric primitives of Gates and LowGates are mapped to to Qualtran `Bloqs,’ being sequences of operations executed on a quantum register, with a parameterized $n$ easing scalable resource estimation by eliminating the need to walk a true circuit.
Composing these Bloqs according to the Lean-emitted pattern allows for a numerical resource estimate for the number of phase rotations, Clifford gates, and Toffoli gates.

Upon conversion to Qualtran, we construct the full order-finding algorithm with numerical verification for each subroutine by operator distance for small instances $|U_{\text{bloq}}-U_{\text{classical}}|$, tracking accumulated phase, and gate count alignment upon decomposition to Toffoli gates, Clifford gates, and controlled rotations.
For a numerical T-gate count estimate, we set a global error of $\delta = 0.01$ for the order-finding algorithm, partitioning half ($\delta_{\text{synth}}=0.005$) to rotation angle synthesis.
Logical T-gate count uses Ross-Selinger synthesis~\cite{ross2014optimal}. 
With a per-rotation error allocation $\varepsilon_{\text{rot}}=\delta_{\text{synth}}/N_{\text{rot}}$ for the $N_\text{rot}$ rotations of the whole order-finding circuit, each rotation is charged $3.075\log_2(1/\varepsilon_{\text{rot}})+0.86$ T-gates, the mean Ross-Selinger cost (confirmed with \texttt{pygridsynth}~\cite{yamamoto2026pygridsynth} on sampled angles).
Logical T-gate count therefore has an additional $\log_2(1/\varepsilon_{\text{rot}})$ scaling due to rotation synthesis.

\begin{figure}[!ht]
\centering
\includegraphics[width=0.8\linewidth]{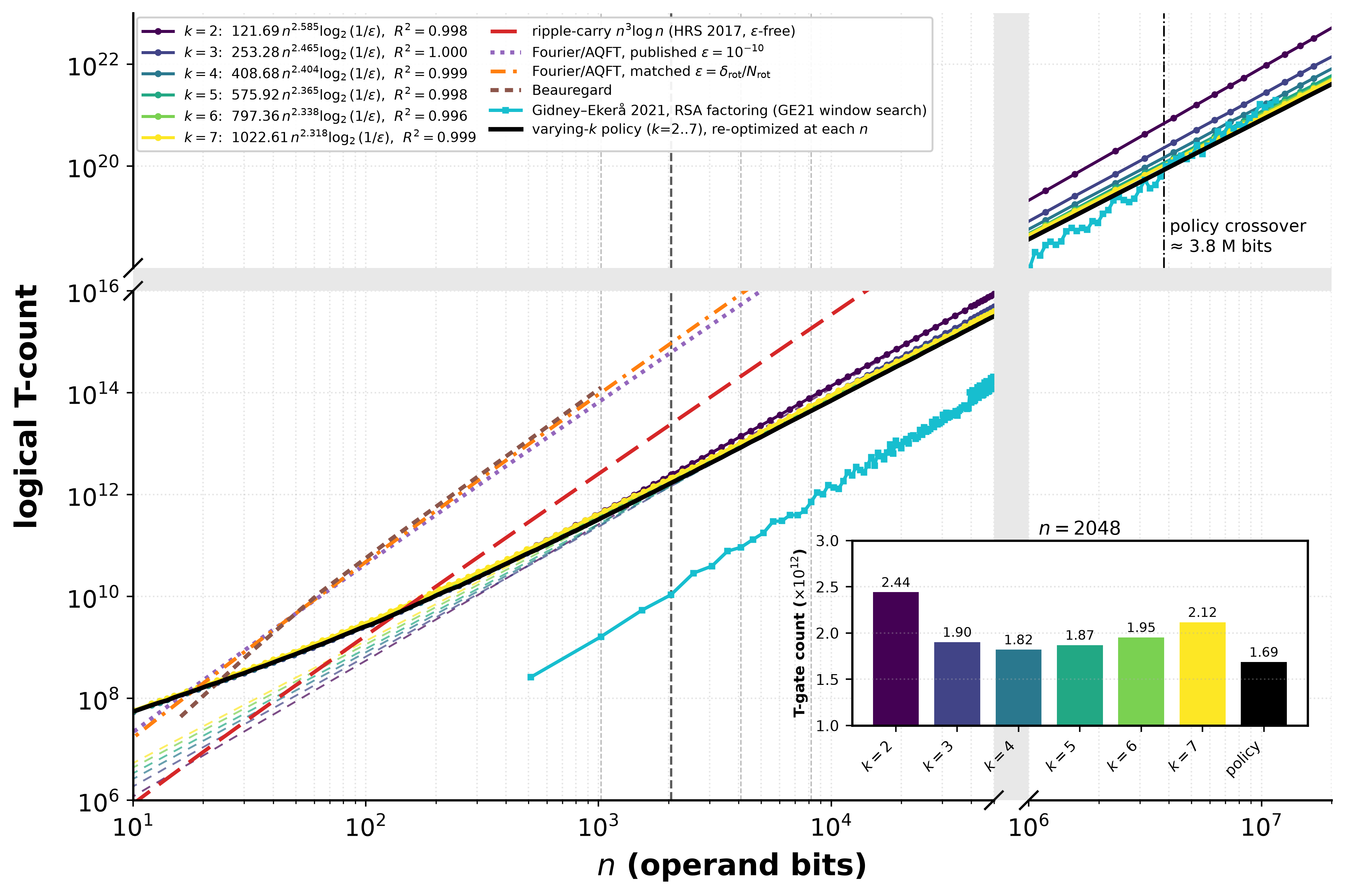}
\caption{Logical T-gate count of Shor's order-finding algorithm using a \texttt{PhaseProduct} construction of Toom-Cook multiplication and QFT.
Following a single $k$-arity brings a best-found T-gate count at $k=4$ with $1.820\times10^{12}$ T-gates, with asymptotic improvements for larger $n$ favoring higher $k$. 
Upon using a varying-$k$ policy for \texttt{PhaseProduct} (different $k$ at different levels in the recursion), a lower T-gate count is identified with $1.685\times 10^{12}$ T-gates.
Least-squares fits assert the derived exponent and fit a constant prefactor to each curve in an asymptotic region $n=[40000,50000]$.
Results are compared to previous instantiations of order-finding, such as Beauregard (2002)~\cite{beauregard2002circuit} (brown dash), H\"aner-Roettler-Svore (2016)~\cite{haner2016factoring} using a ripple-carry addition circuit~\cite{cuccaro2004new, jones2013low} (red dash), the same H\"aner-Roettler-Svore implementation swapping a Fourier addition subroutine~\cite{draper2000addition} with an approximate QFT set at the published error (purple dots), the short discrete logarithm algorithm method of integer factorization of Gidney and Eker\aa{}~\cite{gidney2021factoring}, and our calculated error (orange dot-dash). Without additional optimizations, we find that using \texttt{PhaseProduct} multiplication crosses the short DLP work of Gidney and Eker\aa{} for a modulus size around 3.8 million bits.}\label{fig:TCount}
\end{figure}

For a given $k$ and $n$, the recursion depth is optimized for a minimum T-count, and is compared to a custom policy with varying $k$ at each level in the recursion, reoptimized at each $n$.
At $n=2048$ the main product is decomposed as $k = (6, 5, 3)$ over three recursive layers. 
This policy is not proven optimal and is the focus of future research (Section~\ref{sec:PhasePolicyIsDifficult}).
For constant $k$, we vary integer size $n \in [4, 50000]$, performing a least-squares fit on a range of integer sizes $n=[40000, 50000]$ to represent an `asymptotic region,' fitting $An^{\log_k(2k-1)}\log(1/\varepsilon_\text{rot})$ with some constant $A$, compounding constant factors from additions, multiplications, and synthesis. 
Results are presented in Figure~\ref{fig:TCount}.
The current resource estimation lies well above discrete logarithm methods for integer factorization~\cite{gidney2021factoring}. 
The structure here is fairly straightforward in the substitution of QFT and multiplication routines by incorporating \texttt{PhaseProduct}. 

Significant optimizations are available through compounding optimizations similar to current efforts, such as incorporating other synthesis techniques (fallback~\cite{kliuchnikov2023shorter}, catalysis~\cite{amy2024exact}), zero-ancilla multiplication~\cite{kahanamokuMeyer2024}, approximate QFTs, and physical hardware considerations.
More advanced structural optimizations like windowed multiplication~\cite{gidney2021factoring} may also be applicable with broader implementations, highlighting the improvability of the current resource estimates as formally verified routines evolve.
When compared with another reduction of integer factorization, namely the work of Gidney and Eker\aa{}~\cite{gidney2021factoring} for short discrete logarithms (Figure~\ref{fig:TCount}), we note the T-gate advantage requires an excessively large modulus size of $3.8\times10^6$ bits.
As such, additional optimizations are necessary if the current implementation of Shor's order-finding algorithm is applicable at cryptographically relevant modulus sizes like $n=2048$.
When implementing improved instantiations of Shor's algorithm, a researcher engaging with the challenge can input and compare their work at ForShor.ai.\footnote{\forshorai \, Submitted policies are checked against the harness of Section~\ref{sec:Harness} and priced with the Qualtran backend described here.}.
This online interface allows for dynamic, rapid growth using the \textbf{ForShor} harness with a Qualtran resource estimation backend for quantitative comparison.

\section{Limitations and Future Work}

Although we prove algorithmic correctness and an asymptotic complexity of $O(n^{2+\epsilon})$ in Lean, we do not explicitly prove Lean-to-Qualtran faithfulness, relying on independent numerical validation of each subroutine's action to double precision.
Phase rotations are converted to T-gate estimates by allocating a per-rotation error and charging the mean Ross-Selinger cost $3.075\log_2(1/\varepsilon_{\text{rot}})+0.86$ per rotation (Sec.~\ref{sec:Qualtran}).
This provides an upper-bounded T-gate count estimate, with optimization available with alternative synthesis techniques such as fallback~\cite{kliuchnikov2023shorter} and catalytic embedding~\cite{amy2024exact}.
Numerical resource estimates across modulus sizes $n$ agree with the Lean-proved asymptotic scaling, but are themselves downstream of formal verification.
Further, we leave physical resource estimation to future work, including magic state distillation/cultivation~\cite{litinski2019magic, gidney2024magic}, routing~\cite{constantinides2024optimal, stade2024abstract}, and error-correction~\cite{fowler2012surface, breuckmann2021quantum}. 
The exact optimizer of Appendix~\ref{sec:InterpolationTableOpt} uses a theorem that is functional for free interpolation points, but we limit our implementation to the $K=2$ closed form and, for $K\ge3$, tables built from the KMY points. 
The $n=2048$ compute-time projection rests on measured layer sizes (floors) and a growth law validated at small slack, and is proportional to a time per candidate that we measured once (about $540$~ns at arity $8$, on a shared machine).
We recognize, however, that if the time-per-candidate $\tau$ was even one picosecond (the timescale for a transistor to flip a zero to a one), it would still take an estimated $5.9\times10^{45}$ core hours to exactly enumerate all candidates for $k=8$.

While the work here focuses on Shor's order-finding algorithm, the optimization we focus on is the multiplication subroutine. 
Multiplication has applications across integer factorization algorithms and beyond, including short discrete logarithms~\cite{gidney2021factoring} and Regev's algorithm~\cite{regev2025efficient}.
The latter carries an $O(n^{3/2})$ base gate complexity, requiring $O(\sqrt{n})$ iterations with efficient post-processing, giving a baseline $O(n^2)$ total quantum complexity. 
This is attractive when compared to the $O(n^3)$ base complexity of other integer factorization algorithms and questions the applicability of $O(n^{2+\epsilon})$ order-finding.
Future work aims toward quantitative resource estimation for Regev's algorithm along with the adaptation of PhaseProduct as a modular multiplication subroutine to comprehensively search algorithmic implementations.
 
For a consistent implementation by which T-count can be estimated, we utilize exact QFTs and ripple-carry adders~\cite{cuccaro2004new}. 
Future work targets alternative subroutines to reduce T-count further, such as approximate QFTs~\cite{nam2020approximate} and alternative adders such as Fourier adders~\cite{draper2000addition} or carry-lookahead adders~\cite{thapliyal2021quantum}. 
In practice, approximate operations like approximate QFTs trade significant resource overhead for a small, partitioned error. 
Quantitative resource estimation incorporating these primitives will allow for direct comparison of quantum algorithms collectively aimed toward cryptography, guiding understanding as their utility comes into scope.
Along with a highly-specified subset of valid policies for PhaseProduct requiring all $2k-1$ children have the same working width, the use of a restricted set of interpolation points, picking a table from register width alone, and using the same policy for every modular multiplication instance are various extensions that can further expand the search space and improve T-gate count.
The constructive nature of the \textbf{ForShor} harness allows future research to treat the order-finding architecture as modular.
In this way, future efforts can swap out adders and QFTs, relax the policy constraints imposed currently, and later incorporate synthesis and larger algorithmic improvements while retaining clear validation criteria.
We pose this as one of the greatest strengths of the present framework, currently scoped to exhibit a singular optimization, being fast multiplication.

\section{Conclusion}

\textbf{ForShor} is a Lean~4 harness that formally validates and construct Shor's order-finding algorithm using recursive Toom-Cook \texttt{PhaseProduct} and a Cooley-Tukey QFT, resulting in an algorithm with asymptotic gate complexity of $O(n^{2+\epsilon})$. 
The harness checks validity, preserves circuit semantics, and writes an intermediate representation that can be read into Qualtran for logical resource estimation.
By explicitly defining the algorithmic construction, we allow for contributors to more easily make improvements and optimizations, with formalized details such as finite-width signed arithmetic and padded registers within an operator's recursion lifting difficulties typical of manuscript-level algorithm descriptions.

Formally verifying the quantum circuit, we use an intermediate representation to map its construction in Qualtran. 
At $n=2048$, Ross-Selinger synthesis results in $1.685\times 10^{12}$ T-gates with a custom program with varying $k$ in the recursive definition. 
Toom-Cook $k$-arity varies with modulus size $n$, with direct impact on logical resources following Kahanamoku-Meyer and Yao's asymptotic complexity analysis~\cite{kahanamokuMeyer2024}. 
\textbf{ForShor} focuses on accessibility and optimization of the design space, supporting agentic search for improvement while ensuring fast, formally verified algorithmic correctness. The challenge is open at ForShor.ai.\footref{fn:sites}

\section{Acknowledgement}

 We thank Gregory Kahanamoku-Meyer and Peter Shor for helpful discussions. YC recognizes Unitary Foundation for early support in the project (grant awarded with title ``Quantum factoring resource estimation"), which led to an initial publication \cite{cao2025detailed} describing a Qualtran implementation of \texttt{PhaseProduct} with $k\in\{2,3\}$ and provided a basis for the subsequent collaboration between Zapata Quantum and University of Maryland \cite{zapata2026maryland} that created \textbf{ForShor}.

\printbibliography
 
\appendix
\section{Background and Algorithmic Overview}
\label{sec:Background}
\textbf{ForShor} follows standard textbook constructions for the reduction from factoring to order finding, the quantum order-finding circuit with continued-fraction post-processing, and modular exponentiation through repeated controlled modular multiplications~\cite{shor1997,nielsenChuang2010,beckman1996}. We omit a detailed account of these standard components and focus on the arithmetic and Fourier-transform constructions used to implement the circuit efficiently.

The accompanying website\footnote{\verifiedqc} presents the \textbf{ForShor} development as a graph of Lean declarations, files, and their dependencies. Throughout this appendix, we identify selected graph nodes corresponding to the mathematical objects being introduced, allowing readers to follow the construction alongside its formalization. We begin with \texttt{PhaseProduct}, describe its asymptotic improvement using Toom-Cook arithmetic and the corresponding recursive QFT, and then explain the in-place modular multiplier and the gate complexity of the complete order-finding circuit.

\subsection{PhaseProduct}
\label{Background_PhaseProduct}

The controlled in-place modular multiplication of Section~\ref{Background_ModMul} ultimately relies on a multiplication primitive expressed through phases. This operation, called \texttt{PhaseProduct}, applies a phase proportional to the product of the values stored in two registers. On computational-basis states, it has the form

$$
\ket{x}\ket{z}
\longmapsto
e^{i\phi xz}\ket{x}\ket{z},
$$
for an appropriate phase parameter \(\phi\). 
A direct implementation follows from expanding

$$
xz
=
\sum_{i,j} 2^{i+j}x_i z_j,
$$

where $x_i$ here denotes the $i$-th bit. For each pair of bits \(x_i\) and \(z_j\), the circuit applies a corresponding controlled phase rotation. Thus, for two \(w\)-bit registers, the naive construction uses \(\Theta(w^2)\) pairwise phase operations. On the \textbf{ForShor} proof graph~\cite{forshorWebsite}, this construction can be found by following
    \[
    \texttt{PhaseProduct}
    \;\longrightarrow\;
    \texttt{Gates}
    \;\longrightarrow\;
    \texttt{NaiveLeaf}.
    \]
The \texttt{NaiveLeaf} node contains the direct \texttt{PhaseProduct} construction described above.

This quadratic construction is sufficient for correctness, but using it inside modular multiplication is insufficient for achieving the asymptotic improvement targeted by \textbf{ForShor}. The next subsection describes how Toom-Cook multiplication is adapted to obtain a faster \texttt{PhaseProduct}.

\subsection{Fast \texttt{PhaseProduct} using Toom-Cook Arithmetic}
\label{Background_FastPhaseProduct}

The faster \texttt{PhaseProduct} construction used in \textbf{ForShor} is based on the Toom-Cook multiplication method. We first briefly recall the mathematical idea behind Toom-Cook before describing how it is adapted to \texttt{PhaseProduct}.

\subsubsection{Toom-Cook Multiplication}
\label{Background_ToomCook}

Toom-Cook is a divide-and-conquer method for multiplying large integers~\cite{brentZimmermann2010,bodrato2007}. For a splitting parameter \(k\), each input is divided into \(k\) smaller chunks. These chunks are viewed as the coefficients of two polynomials,

$$
X(t)=\sum_{i=0}^{k-1} x_i t^i,
\qquad
Z(t)=\sum_{i=0}^{k-1} z_i t^i.
$$

Multiplying the original integers can then be reduced to determining the coefficients of the product polynomial
$
C(t)=X(t)Z(t),
$
which has degree at most \(2k-2\).

Instead of directly multiplying every pair of coefficients, Toom-Cook evaluates \(X\) and \(Z\) at \(2k-1\) suitable points, multiplies the corresponding values, and then interpolates these results to recover \(C\)~\cite{brentZimmermann2010,bodrato2007}. Thus, one multiplication problem is reduced to \(2k-1\) smaller multiplication problems, together with the work required for evaluation and interpolation. Recursively applying this idea gives the recurrence

$$
G_k(w)
=
(2k-1)G_k\!\left(\frac{w}{k}\right)+O(w),
$$
and therefore an asymptotic multiplication cost of

$$
G_k(w)
=
O\!\left(w^{\log_k(2k-1)}\right).
$$
The purely mathematical part of this construction can also be followed in the \textbf{ForShor} proof graph~\cite{forshorWebsite}. Starting from the \texttt{PhaseProduct} node, follow

$$
\texttt{PhaseProduct}
\;\longrightarrow\;
\texttt{Math}
\;\longrightarrow\;
\texttt{ToomCook}.
$$

The \texttt{ToomCook} node contains the mathematical definitions and results underlying the splitting, evaluation, and interpolation steps described above. These results are independent of the quantum circuit construction. We now explains how this multiplication strategy is adapted to obtain an asymptotically faster implementation of \texttt{PhaseProduct}.

\subsubsection{Fast \texttt{PhaseProduct}}
\label{Background_FastPhaseProductConstruction}

The Toom-Cook idea described above can be adapted directly to
\texttt{PhaseProduct}, following the construction of KMY~\cite{kahanamokuMeyer2024}.
The evaluation points are the same as in
Toom-Cook, but instead of explicitly reconstructing the product $xz$, the
evaluated values are used to break the phase $\phi xz$ into several smaller
phases.

Split $x$ and $z$ into $k$ chunks of width $h$ and evaluate the chunk polynomials $X$ and $Z$ at $2k-1$ points $p_0,\ldots,p_{2k-2}$. The product $X(t)Z(t)$ has degree at most $2k-2$, and substituting $t=2^h$ gives $xz=\sum_j 2^{jh}c_j$ for its
coefficients $c_j$. The values $X(p_\ell)Z(p_\ell)$ are evaluations of this
polynomial, so $c=(c_0,\dots,c_{2k-2})=V^{-1}\bigl(X(p_\ell)Z(p_\ell)\bigr)_\ell$ for the
interpolation matrix $V$ of the points, and therefore
\[
\phi xz
=
\sum_{\ell=0}^{2k-2}\phi_\ell\,X(p_\ell)\,Z(p_\ell),
\qquad
\phi_\ell=\phi\sum_j 2^{jh}\bigl(V^{-1}\bigr)_{j\ell},
\]
Hence
\[
e^{i\phi xz}
=
\prod_{\ell=0}^{2k-2}
e^{i\phi_\ell X(p_\ell)Z(p_\ell)}.
\]
Each factor on the right is itself a \texttt{PhaseProduct}, on values whose
sizes are roughly a factor of $k$ smaller~\cite{kahanamokuMeyer2024}. The
same construction can therefore be applied recursively. At each level, one
\texttt{PhaseProduct} is replaced by $2k-1$ smaller ones, together with the
reversible additions and shifts required to form the evaluations
$X(p_\ell)$ and $Z(p_\ell)$ in place. Once the registers become sufficiently
small, the recursion stops and the naive \texttt{PhaseProduct} of
Section~\ref{Background_PhaseProduct} is used as the base case.

\begin{table}[t]
    \centering
    \caption{The Toom-Cook interpolation table for $k=2$ with points $0$, $\infty$ and $1$, corresponding to the circuit in Figure~\ref{fig:tc-phaseproduct}. Each arithmetic step is applied to both inputs.}
    \label{tab:tc-k2}
    \renewcommand{\arraystretch}{1.25}
    \setlength{\tabcolsep}{5pt}
    \begin{tabular}{@{} c l c c c @{}}
        \toprule
        \textbf{Step} & \textbf{Operation} & \textbf{Acts on} & \textbf{Point} & \textbf{Coefficient $\phi_{(p)}$} \\
        \midrule
        1 & \texttt{PhaseProduct}                      & $(x_0,\, z_0)$             & $0$      & $\phi\,(1-2^{h})$ \\
        2 & \texttt{PhaseProduct}                      & $(x_1,\, z_1)$             & $\infty$ & $\phi\,(2^{2h}-2^{h})$ \\
        \addlinespace
        3 & Add& $x_0 \leftarrow x_0 + x_1$,\ \ $z_0 \leftarrow z_0 + z_1$ & - & - \\
        4 & \texttt{PhaseProduct}                      & $(x_0+x_1,\, z_0+z_1)$     & $1$      & $\phi\,2^{h}$ \\
        5 & Subtract& $x_0 \leftarrow x_0 - x_1$,\ \ $z_0 \leftarrow z_0 - z_1$ & - & - \\
        \bottomrule
    \end{tabular}
\end{table}

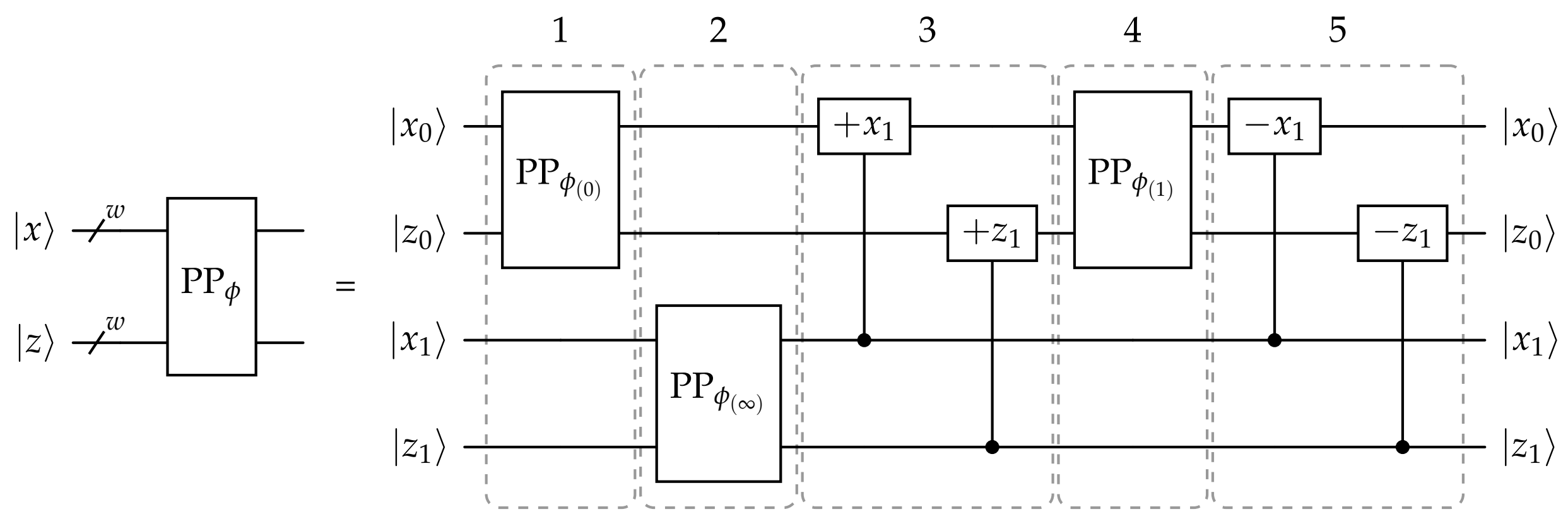
\begin{figure}[t]
    \centering
    \begin{adjustbox}{max width=\linewidth}
    \begin{quantikz}[align equals at=1.5, row sep=0.5cm]
        \lstick{$\ket{x}$} & \qwbundle{w} & \gate[2]{\mathrm{PP}_{\phi}} & \\
        \lstick{$\ket{z}$} & \qwbundle{w} & &
    \end{quantikz}
    \;=\;
    \begin{quantikz}[align equals at=2.5, row sep=0.4cm, column sep=0.4cm]
        \lstick{$\ket{x_0}$} & \gate[2]{\mathrm{PP}_{\phi_{(0)}}}\gategroup[wires=4,steps=1,style={dashed,rounded corners,draw=black!40,inner xsep=1pt},background,label style={label position=above,anchor=south,yshift=-0.1cm}]{1} & \gategroup[wires=4,steps=1,style={dashed,rounded corners,draw=black!40,inner xsep=1pt},background,label style={label position=above,anchor=south,yshift=-0.1cm}]{2} & \gate{+x_1}\gategroup[wires=4,steps=2,style={dashed,rounded corners,draw=black!40,inner xsep=1pt},background,label style={label position=above,anchor=south,yshift=-0.1cm}]{3} & & \gate[2]{\mathrm{PP}_{\phi_{(1)}}}\gategroup[wires=4,steps=1,style={dashed,rounded corners,draw=black!40,inner xsep=1pt},background,label style={label position=above,anchor=south,yshift=-0.1cm}]{4} & \gate{-x_1}\gategroup[wires=4,steps=2,style={dashed,rounded corners,draw=black!40,inner xsep=1pt},background,label style={label position=above,anchor=south,yshift=-0.1cm}]{5} & & \rstick{$\ket{x_0}$} \\
        \lstick{$\ket{z_0}$} & & & & \gate{+z_1} & & & \gate{-z_1} & \rstick{$\ket{z_0}$} \\
        \lstick{$\ket{x_1}$} & & \gate[2]{\mathrm{PP}_{\phi_{(\infty)}}} & \ctrl{-2} & & & \ctrl{-2} & & \rstick{$\ket{x_1}$} \\
        \lstick{$\ket{z_1}$} & & & & \ctrl{-2} & & & \ctrl{-2} & \rstick{$\ket{z_1}$}
    \end{quantikz}
    \end{adjustbox}
    \caption{One level of the Toom-Cook \texttt{PhaseProduct} for $k=2$.
    The numbered boxes are the steps of Table~\ref{tab:tc-k2}, with
    $\phi_{(0)}$, $\phi_{(\infty)}$ and $\phi_{(1)}$ the coefficients listed there.
    Each $\mathrm{PP}$ box is a single \texttt{PhaseProduct} on the two registers
    it spans, and each is itself implemented recursively. The two additions of
    step~3, and the two subtractions of step~5, are drawn one after the other.}
    \label{fig:tc-phaseproduct}
\end{figure}

\paragraph{The case \texorpdfstring{$k=2$}{k=2}.}
Figure~\ref{fig:tc-phaseproduct} illustrates one level of this recursion for
$k=2$, the Karatsuba case. Each $w$-bit input is split into a low chunk of
$h=\lfloor w/2\rfloor$ bits and a high chunk,
\[
x = x_0 + 2^{h}x_1,
\qquad
z = z_0 + 2^{h}z_1,
\]
and the chunk polynomials are evaluated at the three points $0$, $1$ and
$\infty$, giving the pairs $(x_0,z_0)$, $(x_0+x_1,\,z_0+z_1)$ and
$(x_1,z_1)$. A direct calculation shows that
\[
xz
=
(1-2^{h})\,x_0z_0
+ 2^{h}\,(x_0+x_1)(z_0+z_1)
+ (2^{2h}-2^{h})\,x_1z_1,
\]
so the phase $e^{i\phi xz}$ factors into three smaller
\texttt{PhaseProduct}s with coefficients
\[
\phi_{(0)} = \phi\,(1-2^{h}),
\qquad
\phi_{(1)} = \phi\,2^{h},
\qquad
\phi_{(\infty)} = \phi\,(2^{2h}-2^{h}).
\]
The circuit applies these three \texttt{PhaseProduct}s in turn. The products at $0$ and $\infty$ act directly on the chunks $(x_0,z_0)$ and $(x_1,z_1)$. For the product at $1$, the high chunk of each input is first added into its low chunk, the \texttt{PhaseProduct} is applied to the sums, and the additions are then undone, so that every register ends in its original state. The sum $x_0+x_1$ may need one more bit than $x_0$. The figure omits this detail, which is handled by the width analysis.

\paragraph{Interpolation tables.}
Table~\ref{tab:tc-k2} records the $k=2$ construction as a sequence of steps: which pairs of registers receive a \texttt{PhaseProduct}, at which point, and which reversible arithmetic is performed between them. Nothing in this record depends on the width $w$: the steps for $k=2$ are the same whether the inputs have $8$ bits or $8000$. We call such a record an
\emph{interpolation table}. In general an interpolation table for a given $k$
consists of
\begin{itemize}
    \item an ordered list of $2k-1$ evaluation points, each an integer, the
    reciprocal of a nonzero integer, or $\infty$;
    \item a program of shifts, negations and scaled additions on the $k$
          chunk registers, in which each of the $2k-1$ checkpoints marks the
          moment a smaller \texttt{PhaseProduct} is applied to the register
          currently holding the evaluation at the next point, and after
          which the program restores every chunk to its original value.
\end{itemize}
The table does not record the phase coefficients. By the identity above the coefficients
are determined by the interpolation points, computed by the harness. It also does
not record register widths, the extra bits needed to avoid overflow, or the
level at which the recursion stops; Appendix~\ref{sec:AppBEndtoEnd}
explains that these are derived from the table.

A table is therefore a complete specification of the fast
\texttt{PhaseProduct}. Figure \ref{fig:tc-phaseproduct} further highlights this for the $k=2$ example by showing how each step in Table \ref{tab:tc-k2} is mapped onto the \texttt{PhaseProduct} circuit. 

\paragraph{Cost.}
Each of the three smaller \texttt{PhaseProduct}s in the $k=2$ construction
acts on registers of roughly $w/2$ bits and is implemented by the same
construction, until the naive base case is reached. This gives the
recurrence
\[
G_2(w) = 3\,G_2\!\left(\left\lceil\frac{w}{2}\right\rceil + 1\right) + O(w),
\]
and hence a cost of $O(w^{\log_2 3})$. For general $k$, a table with
$2k-1$ points gives the same asymptotic recurrence as Toom-Cook
multiplication,
\[
G_k(w) = (2k-1)\,G_k\!\left(\frac{w}{k}+O(1)\right)+O(w),
\]
and hence a \texttt{PhaseProduct} gate cost of
$G_k(w)=O\!\left(w^{\log_k(2k-1)}\right)$~\cite{kahanamokuMeyer2024}. Since
$\log_k(2k-1)<2$ for $k\geq 2$, this improves asymptotically on the
quadratic naive construction.

\subsection{QFT via Cooley-Tukey Decomposition}
\label{QFTBackground}

The modular-multiplication construction of Section~\ref{Background_ModMul} makes repeated use of the QFT and its inverse. Therefore, accelerating \texttt{PhaseProduct} alone is not sufficient: a quadratic-cost QFT would itself become a bottleneck. Fortunately, the same fast \texttt{PhaseProduct} construction can also be used to obtain a subquadratic exact QFT~\cite{cleve2000fast,kahanamokuMeyer2024}.

\begin{figure}[t]
    \centering
    \begin{adjustbox}{max width=\linewidth}
    \begin{quantikz}[align equals at=1.5, row sep=0.5cm]
        \lstick{$\ket{x_0}$} & \qwbundle{h} & \gate[2]{\mathrm{QFT}_{2^w}} & \rstick{$y_0$} \\
        \lstick{$\ket{x_1}$} & \qwbundle{w-h} & & \rstick{$y_1$}
    \end{quantikz}
    \;=\;
    \begin{quantikz}[align equals at=1.5, row sep=0.5cm, column sep=0.45cm]
        \lstick{$\ket{x_0}$} & \qwbundle{h} & & \gate[2]{\mathrm{PP}_{2\pi/2^w}} & \gate{\mathrm{QFT}_{2^{h}}} & \permute{2,1} & \rstick{$y_0$} \\
        \lstick{$\ket{x_1}$} & \qwbundle{w-h} & \gate{\mathrm{QFT}_{2^{w-h}}} & & & & \rstick{$y_1$}
    \end{quantikz}
    \end{adjustbox}
    \caption{One level of the Cooley--Tukey decomposition of the QFT on a $w$-qubit register split into chunks of $h$ and $w-h$ qubits. The two smaller QFTs are decomposed in the same way, down to single qubits. The procedure is described in the text.}
    \label{fig:ct-qft}
\end{figure}
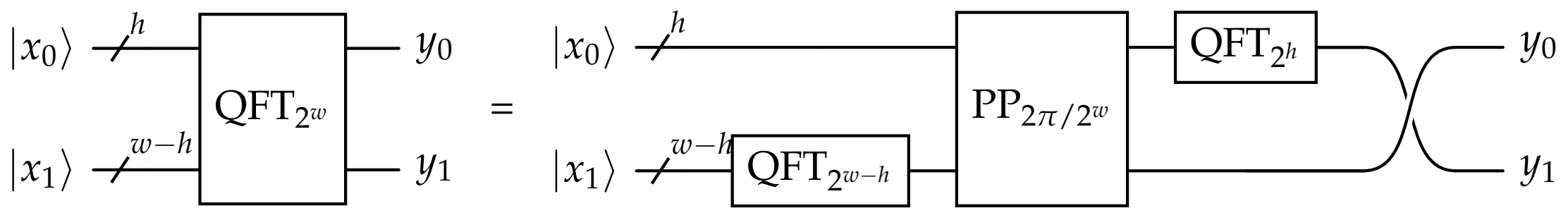

The construction follows a recursive Cooley-Tukey style decomposition of the QFT, shown in Figure~\ref{fig:ct-qft}. As in Section~\ref{Background_FastPhaseProductConstruction}, consider a $w$-qubit register holding $x$, split into a low chunk of $h$ qubits and a high chunk of $w-h$ qubits, so that
\[
x = x_0 + 2^{h}x_1.
\]
The QFT maps $\ket{x}$ to $2^{-w/2}\sum_y e^{2\pi i\,xy/2^w}\ket{y}$. Writing the output in the same way but with the chunk widths exchanged, $y = y_0 + 2^{w-h}y_1$, the phase separates as
\[
\frac{xy}{2^w}
\equiv
\frac{x_1 y_0}{2^{w-h}}
+
\frac{x_0 y_0}{2^{w}}
+
\frac{x_0 y_1}{2^{h}}
\pmod 1,
\]
since the remaining term $x_1 y_1$ is an integer. Each term corresponds to one operation. The first is a QFT on the $(w-h)$-qubit high chunk, which turns $x_1$ into $y_0$. The second is a cross phase between the low chunk and the transformed high chunk, which is exactly the \texttt{PhaseProduct} $\mathrm{PP}_{\phi}$ of Section~\ref{Background_PhaseProduct} with $\phi = 2\pi/2^w$. The third is a QFT on the $h$-qubit low chunk, which turns $x_0$ into $y_1$. Because $y_0$ ends up in the wires that held $x_1$ and vice versa, the two chunks are finally exchanged.

Choosing $h = \lfloor w/2 \rfloor$, as in the $k=2$ split of Section~\ref{Background_FastPhaseProductConstruction}, makes this construction recursive: each of the two smaller QFTs is decomposed in exactly the same way, until the QFTs are small enough to implement directly. Writing $G_k(w)$ for the cost of the Toom-Cook \texttt{PhaseProduct} on $w$-bit registers, the cost of the QFT therefore satisfies
\[
G_{\mathrm{QFT}}(w) = 2\,G_{\mathrm{QFT}}\!\left(\frac{w}{2}\right) + G_k\!\left(\frac{w}{2}\right).
\]
Since $G_k(w) = O\!\left(w^{\log_k(2k-1)}\right)$ and $\log_k(2k-1) > 1$, the \texttt{PhaseProduct} term dominates, and this exact QFT achieves the same asymptotic gate complexity,
\[
G_{\mathrm{QFT}}(w) = O\!\left(w^{\log_k(2k-1)}\right).
\]
\subsection{In-place Modular Multiplication}
\label{Background_ModMul}

We now turn to the in-place controlled modular multiplication required by the modular-exponentiation circuit. For a classical constant $c$ coprime to $N$, the desired operation is
\[
\ket{x}
\longmapsto
\ket{cx\bmod N}.
\]
The fast \texttt{PhaseProduct} developed in the previous subsections is the main multiplication primitive used to construct this operation.

To see how \texttt{PhaseProduct} is used, first consider ordinary multiplication by a classical constant. Let an $m$-qubit register hold $y$, with $\mathrm{QFT}\ket{y}=2^{-m/2}\sum_{z<2^m}e^{2\pi i\,yz/2^m}\ket{z}$. A \texttt{PhaseProduct} with angle $2\pi c/2^m$ between $\ket{x}$ and the transformed register adds $cx$ to $y$:
\begin{align*}
\ket{x}\ket{y}
&\xrightarrow{\;\mathrm{QFT}\;}
\ket{x}\,\frac{1}{\sqrt{2^m}}\sum_{z<2^m}e^{2\pi i\,yz/2^m}\ket{z}\\
&\xrightarrow{\;\mathrm{PP}_{2\pi c/2^m}\;}
\ket{x}\,\frac{1}{\sqrt{2^m}}\sum_{z<2^m}e^{2\pi i\,(y+cx)z/2^m}\ket{z}
=\ket{x}\,\mathrm{QFT}\ket{y+cx\bmod2^m}\\
&\xrightarrow{\;\mathrm{QFT}^{-1}\;}
\ket{x}\ket{y+cx\bmod2^m}.
\end{align*}

The same idea can be adapted to modular multiplication. The key identity is
\[
e^{2\pi i (cx\bmod N)z/N}
=
e^{2\pi i cxz/N}.
\]
The phase therefore automatically discards multiples of $N$: increasing $cx$ by $N$ changes the exponent by an integer multiple of $2\pi i$ and hence leaves the phase unchanged. To exploit this identity, the phase would need denominator $N$, which would require a QFT over $\mathbb{Z}_N$. The QFT on $m$ qubits is over $\mathbb{Z}_{2^m}$ instead, so the phases it produces have denominator $2^m$. Running the construction above in that basis therefore makes the sum wrap around at $2^m$ rather than at $N$: it computes $y + cx \bmod 2^m$, not $y + cx \bmod N$. Since $N$ is odd, no choice of $m$ makes $2^m$ a multiple of $N$, so the reduction modulo $N$ has to be done separately. In that basis the construction above adds $cx$ modulo $2^m$, not modulo $N$.
 
\begin{figure}[t]
    \centering
    \begingroup
    \tikzset{
        modmul step/.style={
            dashed, rounded corners,
            inner xsep=3pt, inner ysep=6pt
        },
        modmul step label/.style={
            label position=above, anchor=south, yshift=1pt
        }
    }
    \begin{adjustbox}{max width=\linewidth}
    \begin{quantikz}[row sep=0.55cm, column sep=0.25cm]
        \lstick{$\ket{0}_{\mathrm{cmp}}$} &
        &[0.5cm] \gategroup[wires=4,steps=3,style={modmul step},background,
            label style={modmul step label}]{Step 1}
        & &
        &[0.5cm] \gategroup[wires=4,steps=3,style={modmul step},background,
            label style={modmul step label}]{Step 2}
        & &
        &[0.5cm] \gate[3]{\mathrm{Cmp}_{\geq N}}
            \gategroup[wires=4,steps=2,style={modmul step},background,
            label style={modmul step label}]{Step 3}
        & \ctrl{1}
        &[0.5cm] \gate[4]{\mathrm{Cmp}_{< Nf}}
            \gategroup[wires=4,steps=1,style={modmul step},background,
            label style={modmul step label}]{Step 4}
        &[0.5cm] \gategroup[wires=4,steps=3,style={modmul step},background,
            label style={modmul step label}]{Step 5}
        & &
        &[0.5cm] \rstick{$\ket{0}_{\mathrm{cmp}}$} \\
        \lstick{$\ket{0}_{o}$} & & & &
        & \gate[2]{\mathrm{QFT}}
        & \gate[3]{\mathrm{PP}_{\alpha_2}}
        & \gate[2]{\mathrm{QFT}^{-1}}
        & & \gate[2]{-N}
        & & & &
        & \rstick{$\ket{0}_{o}$} \\
        \lstick{$\ket{x}$} & \qwbundle{n}
        & & \gate[2]{\mathrm{PP}_{\alpha_1}} &
        & & & & & & &
        & \gate[2]{\mathrm{PP}_{\alpha_3}} &
        & \rstick{$\ket{cx \bmod N}$} \\
        \lstick{$\ket{0}^{\otimes m}$} & \qwbundle{m}
        & \gate{H^{\otimes m}}
        & & \gate{\mathrm{QFT}^{-1}}
        & & & & & &
        & \gate{\mathrm{QFT}}
        & & \gate{\mathrm{QFT}^{-1}}
        & \rstick{$\ket{0}^{\otimes m}$}
    \end{quantikz}
    \end{adjustbox}
    \endgroup
    \caption{In-place modular multiplication $\ket{x}\mapsto\ket{cx \bmod N}$ (Algorithm~1 of~\cite{kahanamokuMeyer2024}). The registers and steps are described in the text.}
    \label{fig:modmul}
\end{figure}
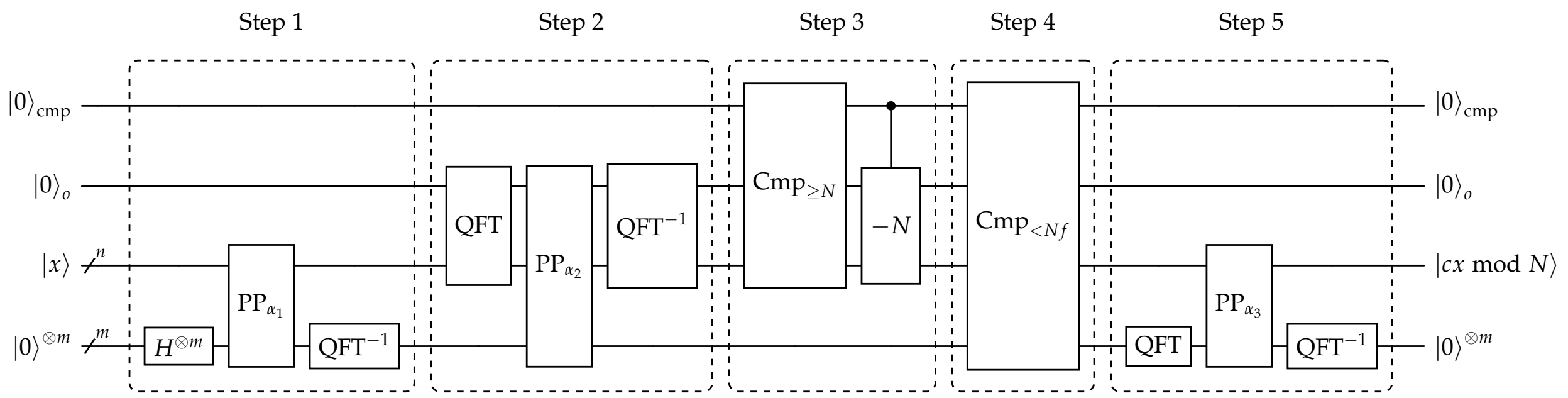

KMY uses these observation to construct an in-place modular multiplier from a sequence of QFTs, \texttt{PhaseProduct}s, and reversible arithmetic operations~\cite{kahanamokuMeyer2024}, and this circuit construction is shown in Figure \ref{fig:modmul}. 
% Figure: in-place modular multiplication (Algorithm 1 of Kahanamoku-Meyer and Yao)

The circuit acts on four registers. The input register $x$ holds the $n$-bit value $x$, and the work register $f$ holds  $m = n + O(\log(1/\varepsilon_{\mathrm{mm}}))$ qubits, where $\varepsilon_{\mathrm{mm}}$ is the target error. Two further qubits are used: an overflow qubit $o$, which extends the input register by one bit, and a comparison qubit $\mathrm{cmp}$. Each block $\mathrm{PP}_{\alpha}$ is a \texttt{PhaseProduct} which, together with the QFTs on either side of it, adds $\alpha$ times the value of one register into the other. The constants are
\[
\alpha_1 = \frac{c-1}{N},
\qquad
\alpha_2 = N,
\qquad
\alpha_3 = -\frac{1-c^{-1}}{N}.
\]
where $c^{-1}$ denotes the inverse of $c$ modulo $N$. The circuit proceeds in five steps:
\begin{enumerate}
    \item \emph{Compute the fraction.} The work register is placed in uniform superposition, $\mathrm{PP}_{\alpha_1}$ is applied between $x$ and $f$, and an inverse QFT is applied to $f$. This writes an $m$-bit approximation of $f = ((c-1)x \bmod N)/N$ into the work register.
    \item \emph{Add $Nf$.} On the input register extended by $o$(the carry bit as shown in the figure), a QFT, $\mathrm{PP}_{\alpha_2}$, and an inverse QFT adds $Nf$ to $x$. Since $x + Nf = x + ((c-1)x \bmod N)$, the extended register now holds an approximation of either $cx \bmod N$ or $(cx \bmod N) + N$.
    \item \emph{Reduce modulo $N$.} A comparison writes into $\mathrm{cmp}$ whether the extended register is at least $N$, and $N$ is subtracted controlled on $\mathrm{cmp}$. The register now holds $cx \bmod N$, and $o$ returns to $\ket{0}$.
    \item \emph{Uncompute the comparison qubit.} The subtraction occurred exactly when $cx \bmod N < Nf$, so comparing the input register with $Nf$ returns $\mathrm{cmp}$ to $\ket{0}$.
    \item \emph{Uncompute the work register.} Since the fraction $f$ equals $((1-c^{-1})(cx) \bmod N)/N$, it can be computed from the new contents of the input register. A QFT on $f$, $\mathrm{PP}_{\alpha_3}$, and an inverse QFT subtract this value, returning the work register to $\ket{0}^{\otimes m}$.
\end{enumerate}

Steps~3 and~4 are exact. The only approximation is that the work register holds $f$ to $m$ bits, while $f$ is a fraction with denominator $N$ that $m$ bits generally cannot represent exactly. The resulting error in $Nf$ is at most about $2^{n-m}$, so each extra bit beyond $n$ halves it, and $m = n + O(\log(1/\varepsilon_{\mathrm{mm}}))$ qubits suffice~\cite{kahanamokuMeyer2024}.

The important feature for our purposes is that the multiplication-intensive components of this construction are \texttt{PhaseProduct}s. The naive \texttt{PhaseProduct} would therefore give a quadratic multiplication cost, whereas replacing it with the recursive Toom-Cook \texttt{PhaseProduct} from Section~\ref{Background_FastPhaseProduct} yields a subquadratic modular multiplier. For a fixed splitting parameter $k$, its gate cost scales as
\[
O\left(n^{\log_k(2k-1)}\right).
\]

The controlled version of this can be implemented by using Controlled-PhaseProducts and leaving the arithmetic operations unchanged would do no harm since they uncompute. Thus, there is only a small constant overhead to this construction. Modular exponentiation applies one such controlled multiplication for each exponent bit.
Thus, an $O(n)$-bit exponent requires $O(n)$ controlled modular multiplications. The subquadratic improvement obtained for \texttt{PhaseProduct} therefore propagates through modular multiplication and ultimately to the complete modular-exponentiation circuit.

The same construction can be followed in the \textbf{ForShor} graph~\cite{forshorWebsite} by opening the top-level \texttt{ModularExponentiation} node. Within this node, the \texttt{Circuit} node contains the concrete implementation of Algorithm~1 from~\cite{kahanamokuMeyer2024}. The \texttt{Proofs} node contains the formal verification of this construction, including several intermediate lemmas establishing the correctness of the individual steps of Algorithm~1. Within \texttt{Proofs}, the \texttt{FinalModMul} node contains the final approximation theorem about the In-Place Controlled Modular Multiplication, which bounds the error between the implemented finite-precision modular multiplication circuit and the corresponding ideal modular multiplication.

\subsection{Order Finding with \texorpdfstring{$O(n^{2+\epsilon})$}{O(n2+eps)} gates}
\label{subsec: Background_gatecount_complexity}
The components of the previous subsections combine as follows. For a fixed splitting parameter $k$, the fast \texttt{PhaseProduct} and the recursive QFT each cost $O(n^{\log_k(2k-1)})$ gates, the in-place controlled modular multiplier inherits the same bound, and modular exponentiation applies $O(n)$ such multiplications, giving $O(n^{1+\log_k(2k-1)})$ for the complete order-finding circuit. Choosing $k$ large enough that $\log_k(2k-1)\leq 1+\epsilon$ gives $O(n^{2+\epsilon})$. Section~\ref{subsec:Guarantees} carries this argument out over the harness's gate-count function on the lowered circuit.

The final construction and its proof can be followed in the \textbf{ForShor} proof graph~\cite{forshorWebsite}. The top-level \texttt{Shor} node contains the complete order-finding circuit and its correctness proof. The resource analysis can be followed through the \texttt{Gate\_Count} node. Within it, \texttt{Shor\_GateCount} contains the final whole-circuit asymptotic argument establishing the $O(n^{2+\epsilon})$ gate bound.

\section{Complexity of the Policy Search}\label{sec:InterpolationTableOpt}

Here we organize the policy search around a classical algorithm that can guarantee an exact, T-optimal policy within the problem's constraints, alongside an estimate for the cost of running said algorithm for $n=2048$.
A recursive \texttt{PhaseProduct} may choose a different Toom-Cook table at every call, where each choice of table changes the arithmetic and the width of lower recursions. 
The cost of synthesizing the leaf rotations can then be calculated for an entire tree search. 
The algorithm below optimizes over \emph{independent} choices at every node, over all finite, legal table programs, and over integer and reciprocal-integer interpolation points (Theorem~\ref{cor:recExact}). 
The exponential factor resides in constructing the cheapest table at a given width, while everything else scales polynomially in $n$ (Section~\ref{sec:AppBInnerOuter}). 

\subsection{Preface: A human summary}\label{sec:AppBPreface}

A \texttt{PhaseProduct} on $w$-bit registers is split into $k$ chunks and evaluated at $2k-1$ interpolation points, each evaluation a smaller \texttt{PhaseProduct}.
A \emph{table} is the order of additions, negations and shifts that forms those evaluations in $k$ register, restoring their valuesa the end.
Each addition is charged once at the child width. In a standalone T-gate optimization, shifts are free but widen registers, so a cheaper table can make every descendant more expensive.
The recursion therefore couples its levels, with rotation synthesis cost of the leaves coupling the whole tree.

\begin{figure}[t]
\centering
\includegraphics[width=0.92\linewidth]{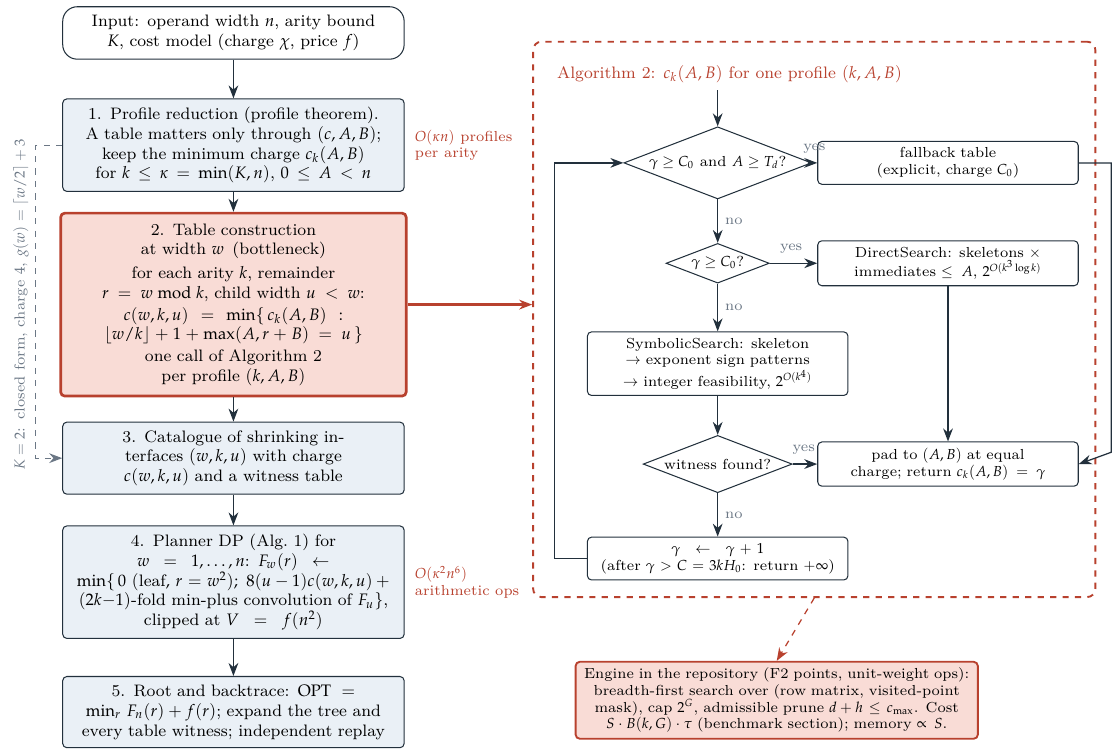}
\caption{Flowchart of the exact planner (Theorem~\ref{cor:recExact}). Step~2, the construction of the cheapest table at each width, is the bottleneck: one call of Algorithm~\ref{alg:profile} per profile $(k,A,B)$ (right). The outer dynamic program (steps 3--5) is polynomial in $n$. The status is \texttt{OPTIMAL} only if the independent replay passes and every arity $\le K$ has a complete catalog (today arity~$2$); otherwise \texttt{catalog\_OPTIMAL}. A timeout or memory exhaustion is reported as incomplete, never as ``no table exists''.}
\label{fig:plannerFlow}
\end{figure}

Here, we build an algorithm that decides at each call whether to stop the recursion, or to choose a table over all finite legal tables.
An \emph{outer} layer is a dynamic program over register widths and leaf-rotation count, combining the children of every call and applying the synthesis cost once. 
This dynamic program is polynomial in $n$ and runs in seconds at $n=2048$ once the tables are known.
The \emph{inner} layer constructs those tables for each arity $k$ and width $w$, finding the cheapest table by exhaustive search in time complexity $2^{O(K^4)}$ for a maximum, manually set arity $K$.
We benchmark this inner layer, with a flowchart of its process in Figure~\ref{fig:plannerFlow}.

We wanted to estimate how long it will take to make a T-gate optimal policy using the algorithm construction given below. 
As such, we focus on the number of states the search visits, the number of candidate moves per state, and the time per candidate $\tau$.
We model or exactly calculate the number of states visited and the number of candidate moves per state, converted to core-hours by $\tau$, which we note is both hardware and software dependent. 
Here, we run GNU Fortran 16.2.0, \texttt{-O3 -std=f95} on an Apple M5 Pro (18 logical CPUs, 24~GiB, macOS 27.2).

We measured $\tau=41$~ns at arity 4, where the tables fit in cache, and \textbf{$\tau=540$~ns at arity 8}, in one run on a shared machine under memory pressure (wall seconds times shards, divided by the candidate evaluations).
Using $540$~ns, we find that for a maximum arity $K$, the memory and time floors and the estimates for certifying the best-known $n=2048$ tree are: $K=8$ needs at least $7.5\times10^{3}$~GiB for any search that stores its states, and an estimated $3.2\times10^{51}$ core-hours (sensitivity span $3.2\times10^{51}$ to $4.9\times10^{80}$); $K=4$ needs an estimated $3.2\times10^{12}$ core hours (span $2.1\times10^{11}$ to $8.8\times10^{15}$).

We do not implement the optimization for freely chosen interpolation points, only covering Appendix F.2 of Kahanamoku-Meyer and Yao~\cite{kahanamokuMeyer2024}.
The Fortran~95 implementation covers the $K=2$ closed form and, for $K\ge3$, catalogs of tables built from the KMY points, with a projection for unit-weight operations.
This preface was written by a human, and the subsequent sections with AI assistance and edited for clarity.
We give proofs for the construction along with a Fortran~95 benchmark below, but have not formally verified the proofs themselves. 
The resulting policies from the Fortran~95 constructions pass the conditions presented in the main text, ensuring validity.

\subsection{Problem Restatement}~\label{sec:AppBNotation}
\subsubsection{Trees and tables}
At each ``node,'' we will have a register width of $w$, which will either be evaluated by a ``leaf,'' being the schoolbook multiplication method, or it will split according to an arity $k$ into $2k-1$ children with width $u<w$, and will have a table ($2\le k\le K$) associated with it. Each child continues independently, so siblings and equal widths do not need to agree.
Fix $q=k-1$, $Q=2k-1$.
A table acts on $k$ rows $X_0,\dots,X_q$, initially the rows of $I_k$ (integer coefficient vectors) and must return them to $I_k$.
Its instructions are $X_i\!\leftarrow\!2^{e}X_i$, $X_i\!\leftarrow\!X_i/2^{e}$ (exact over $\mathbb Z$), $X_i\!\leftarrow\!-X_i$, $X_i\!\leftarrow\!X_i\pm2^{e}X_j$ ($i\ne j$).
It checkpoints each of $Q$ projectively distinct tokens (integers, reciprocals of integers, $0$, $\infty$) exactly once, in the literal evaluation row $v_q(p,s)=(s^q,ps^{q-1},\dots,p^q)$, where sign and scale matter and are not free. 

\subsubsection{Charge and width}\label{app:charge-width}
The \emph{charge} is $\chi=A_++3A_-+\nu$ (additions, subtractions as negate-add-negate, negations); shifts and checkpoints are free.
For $w=kL+r$, $0\le r<k$, the counters $h_i$ start at $(L,\dots,L,L+r)$ and update as $h_i\!+\!e$ (shift), $\max(0,h_i\!-\!e)$ (division), $h_i\!+\!1$ (negation), $1+\max(h_i,h_j\!+\!e)$ (addition); the \emph{child width} is $W_T(w)=1+\max$ of every counter at every instruction boundary, initial and final included.

\subsubsection{Cost}
An internal node $v$ with table $P_v$ and child width $u_v$ pays $8(u_v-1)\chi(P_v)$ (two operand registers, four T gates per adder); a leaf of width $w$ contributes $w^2$ rotation units.
With $D(\mathcal T)=\sum_v8(u_v-1)\chi(P_v)$, $R(\mathcal T)=\sum_{\text{leaves}}w_\ell^2$, $\tau_{\mathrm{rot}}(R)=\operatorname{round}(3.075\log_2(100R)+0.86)$ (evaluated in exact integer arithmetic) and $f(R)=R[4+\tau_{\mathrm{rot}}(R)]$,
\begin{equation}\label{eq:OPT}
\text{opt}(n,K)=\min_{\mathcal T}\ D(\mathcal T)+f\bigl(R(\mathcal T)\bigr).
\end{equation}
For a shared-table policy (one table per arity, uniform levels) this is
\begin{equation}\label{eq:TPP}
\mathsf T_{\mathrm{PP}}=\sum_{i<D}8M_i(w_{i+1}-1)\chi_{k_i}+\mathrm{leaf}(M_D,w_D),\qquad M_{i+1}=M_i(2k_i-1),\quad\mathrm{leaf}(M,w)=f(Mw^2).
\end{equation}
The constants are: $8=4\cdot2$ (four T gates per And, two operand registers); $u-1$ Ands per $u$-bit adder; charges $1,3,1$ (subtraction is negate-add-negate); the Ross--Selinger rotation cost~\cite{ross2014optimal} at $\varepsilon=\delta/2R$, $\delta=0.005$, which gives the $100$ and the exact integer form $\tau_{\mathrm{rot}}(R)=\lfloor(\lfloor\log_2(100R)^{615}\rfloor+272)/200\rfloor$.
Using the width-40 tree as an example, pricing its three leaves separately instead of once gives $84{,}815$ in place of the correct $92{,}750$.

\subsection{Problem Structure and Hardness Context}~\label{sec:AppBStruct}\label{sec:AppBSuppliedPoints}\label{sec:AppBUnproved}
With tables choosing their own points, $\mathrm{OPT}(k)\le15k(2k-1)$, so the language $\mathcal L_S$ is sparse and NP-hardness under many-one reductions would give $\mathrm P=\mathrm{NP}$ (Mahaney~\cite{mahaney1982sparse}; \texttt{Mahaney.tableOpt\_barrier}).
With the points supplied as inputs, the structural problem is NP-hard at $k=2$, by $\mathrm{SAT}\le_p\mathrm{DHC}_2\le_p\mathrm{EHC}\le_p\textsc{PointsTableOpt}$ (\texttt{Pts.genLang\_NPHard}); its hard instances use point sets the benchmark does not admit.
For the T-count, stopping at the root costs $f(n^2)=O(n^2\log n)$, so the unary decision language is sparse and NP-hardness would imply $\mathrm P=\mathrm{NP}$. 
Nowhere in this appendix do we prove T-count optimization is NP-hard, in general or for the KMY points.

\subsection{End-To-End T-gate Count}~\label{sec:AppBEndtoEnd}
\begin{theorem}[Exact free-point independent-tree algorithm]~\label{cor:recExact}
In the model of Section~\ref{sec:AppBNotation}, and given the Lenstra--Kannan integer-feasibility bound~\eqref{eq:LK} for the solver of Algorithm~\ref{alg:profile}, Algorithms~\ref{alg:profile} and~\ref{alg:planner} return $\text{opt}(n,K)$ and an expanded optimal tree with a witness table at every internal node, in
\[O\bigl(|K|+2^{O(\kappa^{4})}\operatorname{poly}(n)\bigr)\ \text{bit operations},\qquad\kappa=\min(K,n).\]
A conservative form is $O(|K|+2^{O(\kappa^4)}(n+2)^{2h+16})$ time and $2^{O(\kappa^4)}(n+2)^{h+8}$ bits of storage for a fixed constant $h\ge2$. As tested, we use $O(n^2[\kappa^3\log(n+2)+n])$ bits.
\end{theorem}
\begin{remark}[Parameterized, not practical]
The theorem is fixed-parameter tractability in $\kappa$ for numerical width $n$. It is not jointly polynomial in $n$ and $K$, and the factor $2^{O(\kappa^4)}$ comes from exhaustive symbolic synthesis.
The imported bound is that integer feasibility in $v$ variables and binary length $\ell$ is decided in
\begin{equation}\label{eq:LK}
2^{O(v\log(v+2))}(\ell+1)^{\beta}\ \text{bit operations for a fixed constant }\beta
\end{equation}
(Lenstra--Kannan~\cite{kannan1987minkowski}; Goemans--Rothvo\ss, Sec.~3, Thm.~4~\cite{goemans2020polynomiality}).
\end{remark}
The proof has four steps, each using only the previous ones: (1)~\emph{compress} each table to a profile (Theorem~\ref{thm:profile}, Lemma~\ref{prop:ceiling}); (2)~\emph{synthesize} the cheapest table at every profile (Lemma~\ref{lem:synth}, Algorithm~\ref{alg:profile}), the bottleneck; (3)~\emph{plan} by a dynamic program over (width, leaf-unit count) (Theorem~\ref{thm:plannerDP}, Algorithm~\ref{alg:planner}); (4)~\emph{assemble} (Section~\ref{sec:AppBInnerOuter}).

\subsubsection{A table as a 3-tuple}
\begin{theorem}[Two-integer profile]\label{thm:profile}
Every arity-$k$ table has unique integers $(a,b)$, $a\ge0$, $\max(0,a-q)\le b\le a$, with $W_T(kL+r)=L+1+\max(a,r+b)$ for $0\le r\le q$; $a$ and $b+q$ are the historical peaks of the counters started from $(0,\dots,0)$ and $(0,\dots,0,q)$. If $(A,B)\ge(a,b)$ componentwise, prefixing $X_0\!\leftarrow\!2^{A}X_0$, $X_0\!\leftarrow\!X_0/2^{A}$, $X_q\!\leftarrow\!2^{B}X_q$, $X_q\!\leftarrow\!X_q/2^{B}$ yields profile exactly $(A,B)$ at unchanged charge.
\end{theorem}
\noindent The proof uses the norm invariant $0<\|X_i\|_1\le2^{h_i}$: no legal division reaches the clamp at $0$.
Replacing a table by a minimum-charge table with the same $(w,k,u)$ keeps the number and widths of all children, hence every continuation and the count $R$, and weakly lowers $D$; only $c(w,k,u)=\min\{c_k(A,B):L+1+\max(A,r+B)=u\}$ matters.
\begin{lemma}[Charge ceiling]\label{prop:ceiling}
Every attainable profile has a representative of charge at most $C=3kH_0=O(k^3)$, $H_0=2q(2q-1)$.
\end{lemma}
\noindent A potential argument gives $\#\text{charged instructions}\le ka$ and $\chi\le3ka$; for $a>H_0$ the explicit fallback table (charge $C_0=2H_0$, profile $(H_0,H_0-q+1)$) is padded to $(a,b)$.

\subsubsection{The cheapest table at a profile: the bottleneck}
\begin{lemma}[Complete local synthesis]\label{lem:synth}
Algorithm~\ref{alg:profile} returns $c_k(A,B)$ and an attaining table, in $2^{O(k^3\log k)}$ time when the budget is at least $C_0$ and $2^{O(k^4)}\operatorname{poly}(\log(A+k+2))$ below it.
\end{lemma}
\noindent\emph{Why unbounded shifts do not escape the search.}
Normalize every addition to the macro $Y_i\!\leftarrow\!2^{p}Y_i$, $Y_i\!\leftarrow\!Y_i\pm2^{s}Y_j$, $Y_i\!\leftarrow\!Y_i/2^{r}$ with $\min(p,s)=0$ and primitive rows between macros; this preserves charge, checkpoints, restoration and every peak, and under cap $A$ all immediates lie in $[0,A]$.
\emph{High budget} ($\gamma\ge C_0$): if $A\ge\max(a_0,b_0+d)$ the fallback fits; otherwise $A=O(k^2)$ and skeletons $\times$ immediates are enumerated, $(6k^2)^{C+Q+1}(A+1)^{2C}=2^{O(k^3\log k)}$.
\emph{Low budget} ($\gamma<C_0$): a skeleton fixes the charged instructions and leaves $p=O(k^2)$ immediates unknown; every matrix entry is a signed sum of powers of two with affine exponents.
Exponents within a gap $H=2c+3$ are resolved exactly; farther clusters have strictly increasing $2$-adic valuations, so correctness is a finite union of integer systems.
Realizable sign patterns are faces of an arrangement of $N=2^{O(k^2)}$ hyperplanes in dimension $p$, at most $(p+1)[2(N+1)]^p=2^{O(k^4)}$, each decided by integer feasibility in $v=O(k^3)$ variables~\eqref{eq:LK}.

\subsubsection{Planning the recursion}
For $1\le w\le n$ let $F_w(r)$ be the least internal cost at width $w$ with exactly $r$ leaf units; the leaf gives $F_w(w^2)=0$, and for each finite interface $(w,k,u)$, with $G_0(0)=0$, $G_j(r)=\min_{1\le s\le r}G_{j-1}(r-s)+F_u(s)$,
\begin{equation}\label{eq:dp}
F_w(r)\leftarrow\min\{F_w(r),\ 8(u-1)\,c(w,k,u)+G_{Q}(r)\},\qquad \text{opt}(n,K)=\min_{1\le r\le n^2}F_n(r)+f(r).
\end{equation}
Since $f(r)/r\ge4$ is nondecreasing, any tree no costlier than the root leaf has $R\le n^2$ and $D\le V=f(n^2)$, which bounds the arrays.
\begin{theorem}[Planner correctness]\label{thm:plannerDP}
With complete interface minima, Algorithm~\ref{alg:planner} returns the optimum~\eqref{eq:OPT} and an attaining independent-node tree, using $O(\kappa^2n^6)$ arithmetic operations on $O(\log n)$-bit numbers.
\end{theorem}
\noindent Induction on $j$ shows $G_j(r)$ is the least total internal cost of $j$ independent children with total count $r$; induction on $w$ covers every feasible tree, and applying $f$ once at the root is exact.
Dominance pruning of child widths (skip $u$ when a smaller width of the same table has a frontier that dominates it pair by pair) is exact and needs no monotonicity of frontiers.

\subsubsection{The arity-two optimum}
\begin{theorem}[All arity-two profile minima]\label{thm:k2table}
$c_2(a,b)=4$ for $b\ge2$ and $+\infty$ for $b<2$, attained by the five-event table $\texttt{chk}\,0$, $\texttt{chk}\,\infty$, $X_0\!\leftarrow\!X_0+X_1$, $\texttt{chk}\,1$, $X_0\!\leftarrow\!X_0-X_1$, with profile $(2,2)$ and child width $g(w)=\lceil w/2\rceil+3$; a shrinking split exists iff $w\ge8$.
\end{theorem}
\begin{theorem}[Forced optimal $K=2$ tree]\label{thm:k2tree}
Let $\mathcal B=\{1,\dots,39,41,43,45\}$. For every $n$, every optimum stops at a width in $\mathcal B$ and otherwise uses charge four and three children of width $g(w)$.
\end{theorem}
\noindent The result holds inside any exterior tree, despite the global price. For $n=2048$ the widths are $2048\to1027\to517\to262\to134\to70\to38$, so there are $3^6=729$ leaves, $R^\ast=1{,}052{,}676$, $D^\ast=739{,}008$ and $\text{opt}(2048,2)=92{,}321{,}820$ (width 40 splits and 41 does not: $92{,}750<92{,}800$ but $97{,}498\le100{,}960$).

\subsection{Inner and Outer Optimization}~\label{sec:AppBInnerOuter}
The \emph{inner} layer builds the interface minima; the \emph{outer} layer is the dynamic program~\eqref{eq:dp}, $O(\kappa^2n^6)$ operations for any catalog. Both are exact, so the cost of the algorithm is the cost of the inner layer: $2^{O(k^4)}$ per profile in the worst case and, as the next section measures, memory-bound well before cryptographic sizes.

We implemented the planner of Theorem~\ref{cor:recExact} in Fortran~95 (the exact $K=2$ closed form; the dense recurrence~\eqref{eq:dp} for $n\le160$ as a value check; a Pareto planner with exact dominance pruning). Interface minima for $k\ge3$ come from a separate exhaustive engine, so the benchmark has two parts: the planner, which is cheap, and the table construction, which is not.
Table programs are checked three ways: an independent big-integer replay that shares no code with the Fortran, compilation of each distinct table in Lean against the benchmark contract, and pricing of the returned tree by the benchmark's own T-count scorer.
\texttt{STATUS OPTIMAL} is printed only after replay and only if every arity $\le K$ has a catalog declared complete (today only arity~$2$, by Theorem~\ref{thm:k2table}); otherwise the status is \texttt{catalog\_OPTIMAL}, optimal relative to the supplied tables and nothing more.

\subsubsection{Execution protocol}
We define a grid of cases being the union of $60,65,\dots,255$, the squares $8^2,\dots,16^2$ and $2^8,\dots,2^{11}$ (50 widths) $\times\ K=2,\dots,5$, each cell three fresh-process repetitions, plus 12 calibration cases.
A cell is launched unless a forecast exceeds 24~h or 16~GiB (median pairwise log--log slope over three widths of equal effective arity, with an upper envelope; screening only); a gated cell is reported as gated, never as infeasible, and no cutoff is imposed after launch.
Time and peak resident size come from \texttt{wait4} usage on one macOS/arm64 machine (Apple M5 Pro, 24~GiB); a cell is accepted only after independent replay.
A $K=4$ or $5$ cell run with an arity-3 catalog records its \emph{effective} arity, $3$, and is an arity-3 result.

\subsubsection{Planner results}
Table~\ref{tab:TOBench} gives the anchor widths for the planner's results.
The $K=2$ optima equal the closed form and an independent oracle. 
The $K=3$ costs are optimal over all F2 tables of arity $\le3$ (the points $\pm2^e$ and their reciprocals): a relaxation with one fictitious zero-charge table of profile $(14,12)$ gives lower bounds equal to the upper bounds at all 50 widths, and exhaustive decisions cover the region it does not (no table with $a\le4$ at any charge; none with charge $\le9$ and $a\le13$; exhaustively tested, one engine).
The planner's wall time grows $\times61$ from $n=512$ to $1024$ and $\times22$ to $2048$ at $K=3$, below the worst case $n^6$ because of dominance pruning (measured, two ratios are not a law).
These times exclude the offline construction of the catalogs.

\begin{table}[!t]
\centering\small
\caption{Planner on the standard grid (report cost model; median of three fresh processes; one machine). $K\!=\!4,5$ cells equal $K\!=\!3$ (effective arity 3).}
\label{tab:TOBench}
\begin{tabular}{@{}rrrlrrl@{}}
\toprule
$n$ & $K$ & $\text{opt}(n,K)$ & status & wall (s) & peak (MiB) & policy\\
\midrule
512 & 2 & 9,078,996 & optimal & 0.009 & 5 & uniform\\
512 & 3 & 7,067,876 & catalog & 0.18 & 9 & uniform\\
1024 & 2 & 29,008,376 & optimal & 0.21 & 17 & uniform\\
1024 & 3 & 20,631,296 & catalog & 11.0 & 58 & mixed\\
2048 & 2 & 92,321,820 & optimal & 5.6 & 126 & uniform\\
2048 & 3 & 62,297,732 & catalog & 241 & 676 & uniform\\
\bottomrule
\end{tabular}
\end{table}

\subsubsection{Projection: the cost of one table construction}\label{sec:AppBProjection}
Table construction at a cell $(k,w,G,c_{\max})$ asks whether a table with at most $c_{\max}$ unit-weight operations exists when every row has weight $W\le2^{G}$ (child width $u=\lfloor w/k\rfloor+G(w,k)$).
The engine (Algorithm~\ref{alg:decide}) has as state (row matrix, mask of visited points); a move negates a row or adds a shifted row; a state at depth $d$ is kept iff $d+h\le c_{\max}$ for an admissible bound $h\le2k$ with $h(I_k)=2k-2$; define the slack $s=c_{\max}-(2k-2)$.
Let $L_d$ be the states first reached after $d$ operations. Then
\begin{align}
B(k,G)&=k+2k(k-1)(2G+1)\quad\text{candidates per expanded state,}\label{eq:B}\\
|L_1|&=4k(k-1)(G-1)+k-4(k-1)\times w\bmod k,\label{eq:N1}\\
|L_{d+1}|&=|L_d|\,r_1\,\lambda^{d-1}\quad(\text{layer law}),\qquad S\approx\textstyle\sum_{d\le s}|L_d|,\label{eq:layers}\\
\text{time}&=\tau\,S\,B,\qquad\text{memory}=b_k\,S,\quad b_k=2k^2+24\ (\text{int16})\ \text{or}\ 4k^2+24\ (\text{int32})\ \text{bytes.}\label{eq:cost}
\end{align}
Equations~\eqref{eq:B} and~\eqref{eq:N1} are exact (\eqref{eq:N1} matches 219 of 219 measured cells, as measured; the layer law~\eqref{eq:layers} has $r_1/|L_1|\in[0.57,0.59]$ and $\lambda\in[0.70,0.75]$ over $k=4,\dots,8$ (measured).
\textbf{The time per candidate $\tau$ is a necessary factor} of every compute-time figure here: $S$ and $B$ are counts, and only $\tau$ turns them into core-hours.
Even though we test a narrow range of values here, we can therefore model $S$ and $B$ and leave the measurement of $\tau$ as the only hardware- and softwaredependent parameter.
We measured $\tau=41$~ns at $k=4$ (best of three, tables in cache) and $\tau=540$~ns at $k=8$: the layer-3 run took $19{,}487$~s on two shards for $(1+|L_1|+|L_2|)\,B=7.3\times10^{10}$ candidates (measured, one run, wall time on a shared machine under memory pressure, so an upper-side value).
Because $\tau$ grew from $41$ to $540$~ns as the working set (and arity) grew, the $41$~ns used for the large arity-4 search is probably optimistic: at $540$~ns the arity-4 estimate would be $4.2\times10^{13}$ core-hours.
The law was tested blind, where before any run we wrote down predicted state counts for three cells not used in the fit, and the exhaustive searches returned $1.564\times10^{7}$ ($1.607\times10^{7}$ predicted), $8.285\times10^{7}$ ($8.452\times10^{7}$) and $9.194\times10^{7}$ ($1.011\times10^{8}$), all within a factor $1.10$ (measured; slack 4--5 only).

\paragraph{What is a bound.}
Because $h\le2k$, every state at depth $d\le c_{\max}-2k$ survives the prune, so any exact state-space search must generate $L_1,\dots,L_3$ for every cell with $c_{\max}\ge2k+3$, and larger caps only add legal moves. Hence at least $B\,|L_2|$ candidate evaluations and $b_k|L_3|$ bytes if states are stored. Table~\ref{tab:layer3} gives the exact $|L_3|$: the memory floor exceeds the 16~GiB gate from $k=6$ and reaches $7.5\times10^{3}$~GiB at $k=8$ (cap $G=23$; $7.3\times10^{10}$ candidate evaluations; the run took $1.9\times10^{4}$~s on two shards, $10.8$ core-hours, under heavy memory pressure). These are measured floors \emph{for state-storing search}, and we do not extrapolate for layers beyond 3.

\begin{table}[!t]
\centering\small
\caption{Measured size of the third layer (states first reached after three operations, smallest cap measured per arity; exact counts).}
\label{tab:layer3}
\begin{tabular}{@{}lrrrrr@{}}
\toprule
arity $k$ & 4 & 5 & 6 & 7 & 8\\
cell $(k,w,G)$ & (0,5) & (0,9) & (0,12) & (0,19) & (0,23)\\
$|L_3|$ & $1.7\!\times\!10^{6}$ & $6.8\!\times\!10^{7}$ & $5.8\!\times\!10^{8}$ & $6.8\!\times\!10^{9}$ & $2.9\!\times\!10^{10}$\\
memory (GiB) & 0.09 & 4.7 & 52 & $1.4\!\times\!10^{3}$ & $7.5\!\times\!10^{3}$\\
\bottomrule
\end{tabular}
\end{table}

\subsubsection{Projection to RSA-2048: floors, estimates and how they are extracted}
\emph{Scope.} An exact optimizer built on this search must decide at least the cells where a relaxation leaves a gap. 
To estimate this task, we show that an artifically constructed best-known tree (arity ladder $8\to4\to3$, board cost $9{,}592{,}816$~T, measured by the board's scorer) is optimal among all F2 policies. 
This ``best-known tree" at $n=2048$ is ``artifical'' in that it takes the charge to be zero i.e. there are no additions or negations, so the dynamic program strictly finds the minimal number of leaf rotations. 
Therefore, this tree is impossible to achieve, but gives a loose lower-bound on T-gate cost. 
Without satisfying conditions C1--C4 as given in the main text, such a policy fails formal verification.
The charge is unit weight (every add or negate is one op); the report's weights give charge in $[c,3c]$, and free (non-F2) points enlarge the state space, so neither is covered.
The extraction is Algorithm~\ref{alg:project}: it prices the best-known tree, finds by free-table relaxation which arities and caps can matter, turns each remaining gap into a cell $(k,w,G_{\mathrm{top}},c_{\max}=c-1)$, closes cells already below a proven bound, and evaluates~\eqref{eq:B}--\eqref{eq:cost} for the rest.
Arity~3 closes (no table with $\le7$ ops at any cap $\le15$, every $w\bmod k$; exhaustive search; arity~4 has the proven bound $c\ge10$ at caps $G_{\mathrm{lb}}..G_{\mathrm{lb}}+3$.
The optimizer is certified optimal over all F2 tables for every benchmark width up to $n=81$; above that the cost steps up with the arity needed.

\begin{table}[!t]
\centering\small
\caption{Certifying the best-known RSA-2048 tree (unit weight, F2 points). Floors are measured and design-relative; estimates are extrapolated; the span is a sensitivity span, \emph{not} a confidence interval.}
\label{tab:T2048}
\begin{tabularx}{\linewidth}{@{}>{\raggedright\arraybackslash}p{0.12\linewidth}>{\raggedright\arraybackslash}p{0.2\linewidth}>{\raggedright\arraybackslash}X>{\raggedright\arraybackslash}p{0.17\linewidth}>{\raggedright\arraybackslash}p{0.2\linewidth}@{}}
\toprule
Arity & Decision (dominant cell) & Floor (measured) & Estimate (core-h) & Lower--upper span\\
\midrule
3 & none $\le7$ ops, $G\le15$ & closed (exhaustive) & $0$ & --\\
4 & none $<20$ ops, $G\le7$ & $|L_3|=1.7\!\times\!10^{6}$ (0.09 GiB) & $3.2\!\times\!10^{12}$ & $2.1\!\times\!10^{11}$--$8.8\!\times\!10^{15}$\\
8 & none $<110$ ops, $G\le31$ & $\ge7.3\!\times\!10^{10}$ candidates ($10.8$ core-h at $540$~ns); $\ge7.5\!\times\!10^{3}$ GiB & $3.2\!\times\!10^{51}$ & $3.2\!\times\!10^{51}$--$4.9\!\times\!10^{80}$\\
\bottomrule
\end{tabularx}
\end{table}

\paragraph{Reading Table~\ref{tab:T2048}.}
Arity~8 dominates, requiring about $3.2\times10^{51}$ core-hours at $\tau=540$~ns ($\approx3.7\times10^{47}$ core-years) against $3.2\times10^{12}$ for arity~4. 
Its state count is the peak of the layer law, $\log_{10}|L|_{\max}\approx(\log_{10}r_1)^2/(2|\log_{10}\lambda|)$, so the exponent scales as $1/|\log\lambda|$.
The span's lower end uses the smallest measured $(\lambda,r_1/|L_1|)$ over arities 5--8 (the $k=8$ cell itself: $\lambda=0.745$, $r_1/|L_1|=0.566$); its upper end the largest, with $\lambda$ multiplied by the largest depth-drift factor measured on exact unpruned layers ($1.10$, at $k=3$). The arity-4-fit scenario (linear-in-$G$ fits at $k=4$ extended to $G=31$) gives $1.1\times10^{64}$.
No coverage probability is claimed: $\lambda$ is measured at one transition for $k\ge5$, its drift only at $k=3$, and the dominant arity-8 cell has slack $109-14=95$ against a calibrated 3--5.
A SAT/CP formulation avoids the memory wall but not the growth: $6.7\times10^{7}$ core-hours for arity~4 at the first cap (growth $\times4.7$ per op from four instances, extrapolated).
The conclusion is a floor and a bracket: certifying the best-known tree needs at least $7.5\times10^{3}$~GiB at arity~8 for any state-storing search and, under the measured growth law, between about $3\times10^{51}$ and $5\times10^{80}$ core-hours; the exponent is not pinned, and the time scale is linear in $\tau$.

For perspective, the memory requirement of $7.3$~TiB is high but achievable using a strong computational cluster as of the time of writing. 
The computation time however, even the lower estimate for arity~4, would take $7.464$~wall years with $1024$ parallel nodes and instantaneous I/O, and an impossible amount of time for $k=8$.
We motivate the policy-optimization challenge on these lower bounds, wherein agentically-driven policy optimization will approach the true optimum faster than brute-force enumeration.

\begin{algorithm}[H]
\caption{$\textsc{Planner}(n,K)$: exact independent-tree optimization (Theorem~\ref{thm:plannerDP})}\label{alg:planner}
\footnotesize
\begin{algorithmic}[1]
\Require width $n\ge1$; arity bound $K\ge2$; interface minima $c(w,k,u)$ with witnesses (Algorithm~\ref{alg:profile})
\State $R_{\max}\gets n^2$;\quad $V\gets f(R_{\max})$
\For{$w=1,\dots,n$}
  \State $F_w[1..R_{\max}]\gets+\infty$;\ $F_w(w^2)\gets0$ \Comment{schoolbook leaf}
  \For{each finite $(w,k,u)$ with $2\le k\le\min(K,w)$, $u<w$}
    \State $G_0(0)\gets0$;\ $G_0(r>0)\gets+\infty$
    \For{$j=1,\dots,2k-1$}
      $G_j(r)\gets\min_{1\le s\le r}\{G_{j-1}(r-s)+F_u(s)\}$ \Comment{independent children}
    \EndFor
    \State $F_w(r)\gets\min\{F_w(r),\ 8(u-1)c(w,k,u)+G_{2k-1}(r)\}$, keeping backpointers; discard costs $>V$
  \EndFor
\EndFor
\State $r^\ast\gets\arg\min_r F_n(r)+f(r)$ \Return $F_n(r^\ast)+f(r^\ast)$ and the backtrace (tables and children)
\end{algorithmic}
\end{algorithm}

\begin{algorithm}[H]
\caption{$\textsc{MinCharge}(k,A,B)$: cheapest arity-$k$ table with exact profile $(A,B)$ (Lemma~\ref{lem:synth})}\label{alg:profile}
\footnotesize
\begin{algorithmic}[1]
\State $q\gets k-1$;\ $d\gets A-B$;\ $H_0\gets2q(2q-1)$;\ $C_0\gets2H_0$;\ $C\gets3kH_0$;\ $T_d\gets H_0+\max(0,d-q+1)$
\For{$\gamma=0,\dots,C$}
  \State $P\gets$ fallback table if $\gamma\ge C_0$ and $A\ge T_d$; else \textsc{DirectSearch}$(k,\gamma,A,d)$ if $\gamma\ge C_0$; else \textsc{SymbolicSearch}$(k,\gamma,A,d)$
  \If{$P\ne\text{none}$} replay $P$; pad to $(A,B)$ (Theorem~\ref{thm:profile}); \Return $(\gamma,P)$ \EndIf
\EndFor
\State \Return $+\infty$ \Comment{only after a complete search; a timeout returns INCOMPLETE}
\end{algorithmic}
\end{algorithm}

\begin{algorithm}[H]
\caption{$\textsc{Decide}(k,w,G,c_{\max})$: is there a table with at most $c_{\max}$ unit-weight ops? (repository engine)}\label{alg:decide}
\footnotesize
\begin{algorithmic}[1]
\State layer $L_0\gets\{(I_k,\text{marks of }0,\infty)\}$;\ store states with full-key hashing; $d\gets0$
\While{$L_d\ne\emptyset$ and $d<c_{\max}$ and no goal found}
  \For{each state $(M,\text{mask})\in L_d$; each of $B(k,G)$ moves: negate a row, or $X_i\!\leftarrow\!2^{p}X_i\pm2^{q}X_j$, $\min(p,q)=0$}
    \State child $(M',\text{mask}')$; \textbf{skip} if a row weight $2\max(P,N)+|{\rm top}|2^{w\bmod k}>2^{G}$ (cap)
    \State \textbf{skip} if already stored, or if $d+1+h(M',\text{mask}')>c_{\max}$ \Comment{admissible, $h\le2k$}
    \State store the child in $L_{d+1}$; \textbf{if} $M'=I_k$ and every point is marked \textbf{then} goal found
  \EndFor
  \State $d\gets d+1$;\ write the exact layer size $|L_d|$
\EndWhile
\State \Return FOUND (replay the path) if goal; NOTFOUND only if the graph was exhausted; UNDECIDED if the state limit was reached
\end{algorithmic}
\end{algorithm}

\begin{algorithm}[H]
\caption{$\textsc{Project}(n)$: floors and estimates for certifying the best-known tree}\label{alg:project}
\footnotesize
\begin{algorithmic}[1]
\State $(T^\ast,\text{ladder})\gets\textsc{Planner}$ over the best-known catalog;\ read the arities used
\State \textbf{relevance:} for each $k\le16$ replace its tables by a free table (charge $0$, child width $\ge u_{\mathrm{lb}}(k,w)$, keep $k$ iff it beats $T^\ast$
\State \textbf{cap cut:} $G_{\mathrm{cut}}\gets$ least cap beyond which free tables do not lower the optimum (tested at $n=512,1024,2048$)
\For{each relevant $k$, residue $w\bmod k$, frontier step $(G_{\mathrm{first}},c)$ with $G_{\mathrm{first}}\le G_{\mathrm{cut}}$}
  \State cell $\gets(k,w,G_{\mathrm{top}},c_{\max}=c-1)$,\ $G_{\mathrm{top}}\gets\min(\text{next }G_{\mathrm{first}}-1,G_{\mathrm{cut}})$,\ slack $s\gets c_{\max}-(2k-2)$
  \If{$c_{\max}$ below a proven bound} cost $0$ (closed) \Comment{arity 3: $\ge8$; arity 4: $\ge10$}
  \Else{} floor $\gets(B\,|L_2|,\ b_k|L_3|)$ from the measured prefix; estimate $\gets\tau\,S\,B$, $b_kS$ by~\eqref{eq:B}--\eqref{eq:cost}
  \EndIf
\EndFor
\State sum over cells; report the dominant cell; span from the extremes of the measured $(\lambda,r_1/|L_1|)$ with the measured depth-drift factor; label extrapolated beyond the calibrated slack
\end{algorithmic}
\end{algorithm}

\end{document}